\documentclass[11pt,a4paper,english]{article}
\usepackage[numbers]{natbib}
\usepackage[latin1]{inputenc}
\usepackage[T1]{fontenc} 
\usepackage[english]{babel}
\usepackage{amsmath,amsfonts,dsfont,amsthm,amssymb,amsbsy,graphicx,graphics,pdfpages, geometry}
\usepackage{subfigure}
\usepackage{booktabs}
\usepackage{slashbox}
\usepackage{url}
\usepackage{url}

\numberwithin{equation}{section}
\theoremstyle{plain}
\newtheorem{theorem}{Theorem}[section]
\newtheorem{lemma}[theorem]{Lemma}
\newtheorem{corollary}[theorem]{Corollary}
\theoremstyle{definition}
\newtheorem{definition}[theorem]{Definition}
\newtheorem{example}[theorem]{Example}

\title{\textbf{Multivariate discrete copulas, with applications in probabilistic weather forecasting}}
\author{Roman Schefzik\vspace{0.2 cm}\\ \textit{Heidelberg Institute for Theoretical Studies}\\ \textit{Schloss-Wolfsbrunnenweg 35, 69118 Heidelberg, Germany}\vspace{0.2 cm}\\ \texttt{roman.schefzik@h-its.org}}
\date{}

\begin{document}

\maketitle

\begin{abstract}
In probability and statistics, copulas play important roles theoretically as well as to address a wide 
range of problems in various application areas. We introduce the concept of multivariate 
discrete copulas, discuss their equivalence to stochastic arrays, and prove a multivariate discrete version of Sklar's 
theorem. These results provide the theoretical frame for multivariate statistical methods to postprocess weather forecasts made 
by ensemble systems, including the ensemble copula coupling approach and the Schaake shuffle.
\end{abstract}


%








\section{Introduction}
Originally introduced by \citet{Sklar1959}, copulas \citep{Nelsen2006} play an important role in probability and statistics whenever the modeling of stochastic dependence is required \citep{Joe2014}. A copula is an $L$-variate cumulative distribution function (CDF) with standard uniform univariate marginal CDFs $F_1=\cdots=F_L=F_{\mathcal{U}([0,1])}$, where $L \in \mathbb{N}$, $L \geq 2$. Hence, copulas form the Fr\'{e}chet class $\mathcal{F}(F_1,\ldots,F_L)=\mathcal{F}(F_{\mathcal{U}([0,1])},\ldots,F_{\mathcal{U}([0,1])})$, where in this context, a Fr\'{e}chet class generally refers to a class of multivariate distributions with fixed uni- or multivariate margins \citep{Frechet1951,Joe1997}. As is manifested in the famous Sklar's theorem \citep{Sklar1959}, a copula links a multivariate CDF to its univariate marginal CDFs. The field of copulas has been developing rapidly over the last decades, and copulas have been applied to a wide range of problems in various areas such as climatology, meteorology and hydrology \citep{Moeller&2013,GenestFavre2007,SchoelzelFriederichs2008} or risk management, insurance  and mathematical finance \citep{Cherubini&2004,Embrechts&2003,PfeiferNeslehova2003}. Moreover, copulas are of theoretical interest, due to their appealing mathematical properties \citep{Joe1997,Nelsen2006,Sempi2011}. However, \citet{Mikosch2006} has pointed out that they also reveal some shortcomings and that an uncritical use is problematic. Hence, copulas cannot be seen as a panacea for all problems connected to the modeling of stochastic dependence \citep{DuranteSempi2010}. A debate about the advantages and disadvantages of copulas can be found in the discussion related to Mikosch's paper \citep{Mikosch2006}.
\\
\\
A special type of copulas are the so-called discrete copulas, whose properties have been studied by \citet{Kolesarova&2006}, \citet{Mayor&2005,Mayor&2007} and \citet{Mesiar2005} in the last decade. Related work in the discrete copula setting can be found in \cite{DenuitLambert2005}, \cite{GenestNeslehova2007}, \cite{MesfiouiTajar2005} and \cite{Neslehova2007}. The discussion in these references focuses on the bivariate case, and an explicit treatment of the general multivariate situation is not conducted, even though it is occasionally mentioned that a generalization of many results is possible \citep{GenestNeslehova2007,Neslehova2007}. Recently, \citet{Genest&2014} have addressed the multivariate case by studying the empirical multilinear copula process based on count data. We also explicitly focus on the multivariate setting in this paper and generalize both the notion of discrete copulas employed by \citet{Kolesarova&2006} and \citet{Mayor&2005} and some of their results \citep{Mayor&2007} from the bivariate to the multivariate case. A new emphasis is put on the equivalence of multivariate discrete copulas and stochastic arrays \citep{Csima1970,MarchiTarazaga1979}. This is employed to prove an extension lemma, which in turn enables us to show a multivariate discrete version of Sklar's theorem. Our approach is clearly driven by the goal to develop a theoretical frame specifically tailored to several multivariate ensemble postprocessing methods that are employed in probabilistic weather forecasting. In particular, the ensemble copula coupling technique \citep{Schefzik&2013} and the Schaake shuffle \citep{Clark&2004}, which are postprocessing methods essentially based on ordering notions, can be interpreted within this frame. Hence, this paper bridges the gap between purely theoretical investigations on multivariate discrete copulas on the one hand and topical applications in probabilistic weather forecasting on the other hand.
\\
\\
The remainder of this paper, which is based on the findings of \citet{Schefzik2014}, 
is organized as follows. In Section \ref{basics}, we introduce the multivariate discrete copula concept. We then point out the equivalence of multivariate discrete copulas and stochastic arrays \citep{Csima1970,MarchiTarazaga1979} in Section \ref{dcstar} and continue with a multivariate discrete version of Sklar's theorem in Section \ref{mdcsklar}. Then, Section \ref{applecc} deals with the relationships of the presented results to multivariate statistical postprocessing methods for ensemble weather forecasts, with a focus on the approaches named above. Finally, the paper closes with a discussion in Section \ref{discussion}.

\section{Multivariate discrete copulas}\label{basics}

\noindent First, we extend the notions of bivariate discrete copulas \citep{Mayor&2005,Kolesarova&2006} and bivariate discrete subcopulas \citep{Mayor&2007} to the general multivariate case.
\newline
\newline
\noindent Let $I_{M}:=\left\{0,\frac{1}{M},\frac{2}{M},\ldots,\frac{M-1}{M},1\right\}$ and $I_M^L:=\underbrace{I_M \times \cdots \times I_M}_{L \operatorname{\,\,times}}$, where $M \in \mathbb{N}$, and let $\overline{\mathbb{R}}:=\mathbb{R} \cup \{-\infty,\infty\}$. 
\begin{definition}\label{mdc}
A function $D: I_{M}^{L} \rightarrow [0,1]$ is a discrete copula on $I_M^L$ if it satisfies the following conditions (D1), (D2) and (D3).
\begin{itemize}
\item[(D1)]{$D$ is grounded, in that $D\left(\frac{i_1}{M},\ldots,\frac{i_L}{M}\right)=0$
if  $i_{\ell}=0$ for at least one $\ell \in \{1,\ldots,L\}$.}
\item[(D2)]{$D(1,\ldots,1,\frac{i_{\ell}}{M},1,\ldots,1)=\frac{i_{\ell}}{M}$ for all $\ell \in \{1,\ldots,L\}$.}
\item[(D3)]{$D$ is $L$-increasing, in that
\begin{equation*}
\Delta_{i_{L}-1}^{i_{L}} \cdots \Delta_{i_{1}-1}^{i_{1}} D \left(\frac{k_{1}}{M},\ldots,\frac{k_{L}}{M}\right) \geq 0
\end{equation*} 
for all $i_{\ell} \in \{1,\ldots,M\}$, $\ell \in \{1,\ldots,L\}$, where
\begin{eqnarray*}
\Delta_{i_{\ell}-1}^{i_{\ell}} D \left(\frac{k_{1}}{M},\ldots,\frac{k_{L}}{M}\right)&:=&D\left(\frac{k_{1}}{M},\ldots,\frac{k_{\ell-1}}{M},\frac{i_{\ell}}{M},\frac{k_{\ell+1}}{M},\ldots,\frac{k_{L}}{M}\right)\nonumber\\&&-D\left(\frac{k_{1}}{M},\ldots,\frac{k_{\ell-1}}{M},\frac{i_{\ell}-1}{M},\frac{k_{\ell+1}}{M},\ldots,\frac{k_{L}}{M}\right).
\end{eqnarray*}
}
\end{itemize}
\end{definition}
\begin{definition}\label{irrmdc}
A discrete copula $D:  I_{M}^{L} \rightarrow [0,1]$ is irreducible if it has minimal range, that is, $\mbox{Ran}(D)=I_{M}$.
\end{definition}
\noindent Following \citet{Frechet1951} and \citet{Joe1997}, a multivariate discrete copula can be interpreted as a multivariate distribution in the Fr\'{e}chet class $\mathcal{F}(F_{\mathcal{U}(K)},\ldots,F_{\mathcal{U}(K)})$, where $F_{\mathcal{U}(K)}$ is the CDF of a uniformly distributed random variable on a set $K$ of cardinality $M$.
\begin{definition}\label{mdsubcop}
Let $\{0,1\} \subset J_{M}^{(1)},\ldots, J_{M}^{(L)} \subset I_{M}$. A function $D^{\ast}: J_{M}^{(1)} \times \cdots \times J_{M}^{(L)} \rightarrow [0,1]$ is a discrete subcopula if it satisfies the following conditions (S1), (S2) and (S3).
\begin{itemize}
\item[(S1)]{$D^{\ast}\left(\frac{i_1}{M},\ldots,\frac{i_L}{M}\right)=0$
if $i_{\ell}=0$ for at least one $\ell \in \{1,\ldots,L\}$.}
\item[(S2)]{$D^{\ast}(1,\ldots,1,\frac{i_{\ell}}{M},1,\ldots,1)=\frac{i_{\ell}}{M}$ for all $\frac{i_{\ell}}{M} \in J_{M}^{(\ell)}$.}
\item[(S3)]{
\begin{equation*}
\Delta_{i_{L}}^{j_{L}} \cdots \Delta_{i_{1}}^{j_{1}} D^{\ast} \left(\frac{k_{1}}{M},\ldots,\frac{k_{L}}{M}\right) \geq 0
\end{equation*}
for all $(\frac{i_{1}}{M},\ldots,\frac{i_{L}}{M}), (\frac{j_{1}}{M},\ldots,\frac{j_{L}}{M}) \in J_{M}^{(1)} \times \cdots \times J_{M}^{(L)}$ such that $i_{\ell} \leq j_{\ell}$ for $\ell \in \{1,\ldots,L\}$, where
\begin{eqnarray*}
\Delta_{i_{\ell}}^{j_{\ell}} D^{\ast} \left(\frac{k_{1}}{M},\ldots,\frac{k_{L}}{M}\right)&:=&D^{\ast}\left(\frac{k_{1}}{M},\ldots,\frac{k_{\ell-1}}{M},\frac{j_{\ell}}{M},\frac{k_{\ell+1}}{M},\ldots,\frac{k_{L}}{M}\right)\\&&-D^{\ast}\left(\frac{k_{1}}{M},\ldots,\frac{k_{\ell-1}}{M},\frac{i_{\ell}}{M},\frac{k_{\ell+1}}{M},\ldots,\frac{k_{L}}{M}\right).
\end{eqnarray*}
}
\end{itemize}
\end{definition}
\begin{definition}\label{irredsubcop}
A discrete subcopula $D^{\ast}: J_{M}^{(1)} \times \cdots \times J_{M}^{(L)} \rightarrow [0,1]$ is irreducible if $\operatorname{Ran}(D^\ast)=I_M$. 
\end{definition}
\noindent The definition of discrete (sub)copulas can be generalized. A discrete copula need not necessarily have domain $I_{M}^{L}$, but can generally be defined on $I_{M_{1}} \times \cdots \times I_{M_{L}}$, where $M_{1},\ldots,M_{L} \in \mathbb{N}$ might take distinct values. Then, the axioms (D1), (D2) and (D3) apply analogously to this case. Similarly, discrete subcopulas can generally be defined on $J_{M_{1}}^{(1)} \times \cdots \times J_{M_{L}}^{(L)}$ for possibly distinct numbers $M_{1},\ldots,M_{L} \in \mathbb{N}$, satisfying the conditions in Definition \ref{mdsubcop}. For convenience and in view of the applications in Section \ref{applecc}, we confine ourselves to the case of $M:=M_{1}=\cdots=M_{L}$ as in the above Definitions \ref{mdc} to \ref{irredsubcop} in what follows.
\\
\\
Further tailored to the applications in Section \ref{applecc}, we defined the multivariate discrete copula on points that are equally spaced across the set $I_M^L$. However, when considering a multivariate distribution with discrete margins, the points where the copula is of interest typically do not need to be equidistant across $I_M^L$, but rather are heterogeneously spaced across the margins. Such more general situations are studied in \cite{GenestNeslehova2007} and \cite{Genest&2014}, for instance.  
\begin{example}\label{pim}
Let $i_1,\ldots,i_M \in \{0,1,\ldots,M\}$.
\begin{enumerate}
\item[(a)]{The function
\begin{equation*}
\Pi\left(\frac{i_1}{M},\ldots,\frac{i_L}{M}\right):=\prod\limits_{\ell=1}^{L}\frac{i_{\ell}}{M} 
\end{equation*}
is a discrete copula. It represents the restriction of the so-called product or independence copula from $[0,1]^{L}$ to $I_{M}^{L}$.}
\item[(b)]{The function
\begin{equation*}
{\cal{M}}\left(\frac{i_1}{M},\ldots,\frac{i_L}{M}\right):=\mbox{min}\left\{\frac{i_{1}}{M},\ldots,\frac{i_{L}}{M}\right\} 
\end{equation*}
is an irreducible discrete copula. It represents the restriction of the so-called comonotonicity copula, which is the Fr\'{e}chet-Hoeffding upper bound and models perfect positive dependence, from $[0,1]^{L}$ to $I_{M}^{L}$.}
\item[(c)]{Another example for an irreducible discrete copula  
is given by the so-called empirical copula \citep{Rueschendorf1976}, which has also become popular under the term ``empirical dependence function'' \citep{Deheuvels1979}. The empirical copula is particularly relevant in view of the applications to be discussed in Section \ref{applecc}, for which it provides the theoretical background.
\\
Let  
${\cal{S}}:=\{(x_{1}^{1},\ldots,x_{1}^{L}),\ldots,(x_{M}^{1},\ldots,x_{M}^{L})\}$, where  
$x_{m}^{\ell} \in  \mathbb{R}$ for $m \in \{1,\ldots,M\}$ and $\ell  
\in \{1,\ldots,L\}$ with $x_{m}^{1} \neq x_{\mu}^{1},\ldots,x_{m}^{L} \neq  
x_{\mu}^{L}$ for $m,\mu \in \{1,\ldots,M\}$, $m \neq \mu$. That is, we assume for simplicity that there are no ties among the respective samples. Moreover, let  
$x_{(1)}^{1}< \ldots < x_{(M)}^{1},\ldots,x_{(1)}^{L}<\ldots<x_{(M)}^{L}$ be the  
marginal order statistics of the collections\\  
$\{x_{1}^{1},\ldots,x_{M}^{1}\},\ldots,\{x_{1}^{L},\ldots,x_{M}^{L}\}$,  
respectively.
\\
Then, the empirical copula $E_{M}:I_{M}^{L} \rightarrow I_{M}$ defined  
from ${\cal{S}}$ is given by
\begin{align*}
&E_{M}\left(\frac{i_{1}}{M},\ldots,\frac{i_{L}}{M}\right)\\&:= 
\begin{cases} 0&\mbox{if }i_{\ell}=0 \mbox{\,\,for at least one } \ell \in  
\{1,\ldots,L\}, \\
\frac{\#\{(x_{m}^{1},\ldots,x_{m}^{L}) \in {\cal{S}} | x_{m}^{1} \leq  
x_{(i_{1})}^{1},\ldots,x_{m}^{L} \leq x_{(i_{L})}^{L} \}}{M}&\mbox{if\,  
}i_{\ell} \in \{1,\ldots,M\} \mbox{\,\,for all } \ell \in \{1,\ldots,L\}.
\end{cases}
\end{align*}
Equivalently,
\begin{eqnarray*}
E_{M}\left(\frac{i_{1}}{M},\ldots,\frac{i_{L}}{M}\right) &=&\frac{1}{M} \sum\limits_{m=1}^{M}\mathds{1}_{\{\operatorname{rank}(x_{m}^{1}) \leq i_1,\ldots,\operatorname{rank}(x_{m}^{L})\leq i_L\}}\\ &=& \frac{1}{M}  
\sum\limits_{m=1}^{M} \prod\limits_{\ell=1}^{L}  
\mathds{1}_{\left\{\operatorname{rank}(x_{m}^{\ell}) \leq i_{\ell}\right\}},
\end{eqnarray*}
where $\operatorname{rank}(x_m^\ell)$ denotes the rank of $x_m^\ell$ in $\{x_1^\ell,\ldots,x_M^\ell\}$ for $m \in \{1,\ldots,M\}$ and $\ell \in \{1,\ldots,L\}$ \citep{Deheuvels1979}.
\\
Obviously, the empirical copula is an irreducible discrete copula.  
Conversely, any irreducible discrete copula is the empirical copula of  
some set ${\cal{S}}$, as will be discussed in Example \ref{exequiv} (c) in  
Section \ref{dcstar}.
\\
Asymptotic theory for the corresponding empirical processes is readily available \citep{Rueschendorf1976,VanderVaartWellner1996,Fermanian&2004,Rueschendorf2009}.}
\item[(d)]{An explicit example for a discrete copula in the case of $L=3$ and $M=3$ is given by the function $D$ specified in Table \ref{tab1}. The axioms (D1), (D2) and (D3) from Definition \ref{mdc} can be verified straightforwardly. 
\begin{table}[t]
\caption{\small{Example \ref{pim} (d) of a discrete copula $D$ in the case of $L=3$ and $M=3$: The explicit values $D\left(\frac{i_{1}}{3},\frac{i_{2}}{3},\frac{i_{3}}{3}\right)$ are given.\vspace{0.2 cm}}}\label{tab1}
\centering
\footnotesize
 \begin{minipage}[b]{.49\linewidth}
 \centering
$i_{1}=0$\\
\begin{tabular}{c|cccc}
\toprule
\backslashbox[4mm]{$i_{3}$\kern-1em}{\kern-1em$i_{2}$}&$0$&$1$&$2$&$3$\\ \hline 
$0$&$0$&$0$&$0$&$0$\\
$1$&$0$&$0$&$0$&$0$\\
$2$&$0$&$0$&$0$&$0$\\
$3$&$0$&$0$&$0$&$0$\\
\bottomrule
\end{tabular}
  \end{minipage}
\begin{minipage}[b]{.49\linewidth} 
\centering
$i_{1}=1$\\
\begin{tabular}{c|cccc}
\toprule
\backslashbox[4mm]{$i_{3}$\kern-1em}{\kern-1em$i_{2}$}&$0$&$1$&$2$&$3$\\ \hline 
$0$&$0$&$0$&$0$&$0$\\
$1$&$0$&$1/12$&$1/6$&$1/6$\\
$2$&$0$&$1/12$&$1/6$&$1/4$\\
$3$&$0$&$1/12$&$1/6$&$1/3$\\
\bottomrule
\end{tabular}
  \end{minipage}
   \begin{minipage}[b]{.49\linewidth} 
   \centering
$i_{1}=2$\\
\begin{tabular}{c|cccc}
\toprule
\backslashbox[4mm]{$i_{3}$\kern-1em}{\kern-1em$i_{2}$}&$0$&$1$&$2$&$3$\\ \hline 
$0$&$0$&$0$&$0$&$0$\\
$1$&$0$&$1/12$&$1/4$&$1/4$\\
$2$&$0$&$1/6$&$1/3$&$1/2$\\
$3$&$0$&$1/4$&$5/12$&$2/3$\\
\bottomrule
\end{tabular}
  \end{minipage}
   \begin{minipage}[b]{.49\linewidth} 
   \centering
   \vspace{0.5 cm}
   $i_{1}=3$\\
\begin{tabular}{c|cccc}
\toprule
\backslashbox[4mm]{$i_{3}$\kern-1em}{\kern-1em$i_{2}$}&$0$&$1$&$2$&$3$\\ \hline 
$0$&$0$&$0$&$0$&$0$\\
$1$&$0$&$1/12$&$1/3$&$1/3$\\
$2$&$0$&$1/4$&$1/2$&$2/3$\\
$3$&$0$&$1/3$&$2/3$&$1$\\
\bottomrule
\end{tabular}
  \end{minipage} 
\end{table}
}
\end{enumerate}
\end{example}



\section{A characterization of multivariate discrete copulas using stochastic arrays}\label{dcstar}

\noindent According to \cite{Kolesarova&2006} and \cite{Mayor&2005}, there is a one-to-one correspondence between discrete copulas and bistochastic matrices in the bivariate case. We now formulate a similar characterization for multivariate discrete copulas. To this end, the notion of stochastic arrays \citep{Csima1970,MarchiTarazaga1979} is required.
\begin{definition}\label{stocharr}
An array $A:=(a_{i_{1} \ldots i_{L}})_{i_{1},\ldots,i_{L}=1}^{M}$ is an $L$-dimensional stochastic array, or an $L$-stochastic matrix, of order $M$ if the following conditions (A1) and (A2) hold. 
\begin{itemize}
\item[(A1)]{$a_{i_{1} \ldots i_{L}} \geq 0$\,\, for all $i_{1},\ldots,i_{L} \in \{1,\ldots,M\}$.}
\item[(A2)]{$\sum\limits_{i_{1}=1}^{M} \cdots \sum\limits_{i_{\ell-1}=1}^{M} \sum\limits_{i_{\ell+1}=1}^{M} \cdots \sum\limits_{i_{L}=1}^{M} a_{i_{1} \ldots i_{\ell-1} i_{\ell} i_{\ell+1} \ldots i_{L}}=1$\,\, for\,\, $i_{\ell} \in \{1,\ldots,M\}$,\, $\ell \in \{1,\ldots,L\}$.}
\end{itemize}
\noindent As a special case, an $L$-dimensional stochastic array $A$ is an $L$-dimensional permutation array, or an $L$-permutation matrix, if the entries of $A$ only take the values 0 and 1, that is, $a_{i_{1} \ldots i_{L}} \in \{0,1\}$ for all $i_{1},\ldots,i_{L} \in \{1,\ldots,M\}$.
\end{definition}
\begin{theorem}\label{equiv}
Let $D:I_{M}^{L} \rightarrow [0,1]$. Then, the following statements (1) and (2) are equivalent.
\begin{enumerate}
\item[(1)]{$D$ is a discrete copula.}
\item[(2)]{There exists an $L$-dimensional stochastic array $A:=(a_{i_{1} \ldots i_{L}})_{i_{1},\ldots,i_{L}=1}^{M}$ of order $M$ such that
\begin{equation}\label{stochm}
D \left(\frac{i_{1}}{M},\ldots,\frac{i_{L}}{M}\right)=\frac{1}{M} \sum\limits_{\nu_{1}=1}^{i_{1}} \cdots \sum\limits_{\nu_{L}=1}^{i_{L}} a_{\nu_{1} \ldots \nu_{L}}
\end{equation}
for $i_{1},\ldots,i_{L} \in \{0,1,\ldots,M\}$.
}
\end{enumerate}
\end{theorem}
\noindent A proof of Theorem \ref{equiv} is given in Appendix \ref{equiv.proof}.
\begin{corollary}\label{corequiv}
$D$ is an irreducible discrete copula if and only if there exists an $L$-dimensional permutation array $A:=(a_{i_{1} \ldots i_{L}})_{i_{1},\ldots,i_{L}=1}^{M}$ such that $\eqref{stochm}$ holds for $i_{1},\ldots,i_{L} \in \{0,1,\ldots,M\}$.
\end{corollary}
\noindent Theorem \ref{equiv} can also be interpreted as a reformulation of the relation between the CDF and the probability mass function (PMF) \citep{Xu1996} because the stochastic array in Definition \ref{stocharr} can be identified with $M$ times the PMF.
\\
\\
Essentially, Theorem \ref{equiv} yields the equivalences
\\
\hspace*{4 mm} Discrete copula\\
 $\Leftrightarrow$ Marginal distributions concentrated on $I_M$\\
  $\Leftrightarrow$ Probability masses on $\left\{\frac{1}{M},\frac{2}{M},\ldots,1 \right\}^L$\\
   $\Leftrightarrow$ Stochastic array.
\\
\\
\noindent In the situation of Corollary \ref{corequiv}, we have the equivalences\\
\hspace*{4 mm} Irreducible discrete copula\\
 $\Leftrightarrow$ Empirical copula\\
  $\Leftrightarrow$ $M$ point masses of $\frac{1}{M}$ each\\
   $\Leftrightarrow$ Permutation array\\
    $\Leftrightarrow$ Latin hypercube of order $M$ in $L$ dimensions \citep{Gupta1974}.
\\
\\
Illustrations of these equivalences are given in Section \ref{applecc}, where we discuss their relevance with respect to applications in probabilistic weather forecasting.
\begin{example}\label{exequiv}
$\hfill$
\begin{enumerate}
\item[(a)]{The discrete product copula $\Pi\left( \frac{i_{1}}{M},\ldots,\frac{i_{L}}{M}\right) := \prod\limits_{\ell=1}^{L}\frac{i_{\ell}}{M}$ on $I_{M}^{L}$ from Example \ref{pim} (a) corresponds to the $L$-dimensional stochastic array $A:=\left(\frac{1}{M^{L-1}}\right)_{i_{1},\ldots,i_{L}=1}^{M}$ of order $M$ whose entries are all equal to $\frac{1}{M^{L-1}}$. Indeed,
\begin{eqnarray*}
\frac{1}{M} \sum\limits_{\nu_{1}=1}^{i_{1}} \cdots \sum\limits_{\nu_{L}=1}^{i_{L}}\frac{1}{M^{L-1}} &=& \frac{1}{M} \sum\limits_{\nu_{1}=1}^{i_{1}} \cdots \sum\limits_{\nu_{L-1}=1}^{i_{L-1}}\frac{i_{L}}{M^{L-1}}
=\frac{1}{M} \cdot \frac{i_{1}\cdot \ldots \cdot i_{L-1}\cdot i_{L}}{M^{L-1}}\\
&=& \prod\limits_{\ell=1}^{L}\frac{i_{\ell}}{M}
= \Pi\left(\frac{i_{1}}{M},\ldots,\frac{i_{L}}{M}\right).
\end{eqnarray*}
}
\item[(b)]{The irreducible discrete copula ${\cal{M}}\left( \frac{i_{1}}{M},\ldots,\frac{i_{L}}{M}\right):= \mbox{min}\left\{ \frac{i_{1}}{M},\ldots,\frac{i_{L}}{M}\right\} $ on $I_{M}^{L}$ in Example \ref{pim} (b) corresponds to the $L$-dimensional identity permutation array
\begin{equation*}
\mathbb{I}:=(a_{i_{1} \ldots i_{L}})_{i_{1},\ldots,i_{L}=1}^{M}, \mbox{\,\, where\,\,} a_{i_{1} \ldots i_{L}}:=\begin{cases} 1&\mbox{if\,}i_{1}=\cdots=i_{L} \\ 0& \mbox{otherwise}\end{cases},
\end{equation*}
of order $M$. Indeed, employing the definition and writing down the corresponding multiple sum explicitly yields
\begin{eqnarray*}
\frac{1}{M}\sum\limits_{\nu_{1}=1}^{i_{1}} \cdots \sum\limits_{\nu_{L}=1}^{i_{L}}a_{\nu_{1} \ldots \nu_{L}} &=& \frac{1}{M} \cdot \mbox{min}\{i_{1},\ldots,i_{L}\}= \mbox{min}\left\{\frac{i_{1}}{M},\ldots,\frac{i_{L}}{M}\right\}\\
&=&{\cal{M}}\left(\frac{i_{1}}{M},\ldots,\frac{i_{L}}{M}\right).
\end{eqnarray*}
}
\item[(c)]{The empirical copula $E_{M}$ in Example \ref{pim} (c), being an  
irreducible discrete copula, corresponds to the $L$-dimensional  
permutation array $A:=(a_{i_{1} \ldots i_{L}})_{i_{1},\ldots,i_{L}=1}^{M}$ of order $M$ with
\begin{equation*}
a_{i_{1} \ldots i_{L}}:=\begin{cases} 1&\mbox{if } (x_{(i_{1})}^{1},\ldots,x_{(i_{L})}^{L}) \in  
{\cal{S}}, \\
0&\mbox{if } (x_{(i_{1})}^{1},\ldots,x_{(i_{L})}^{L}) \notin {\cal{S}},
\end{cases}
\end{equation*}
with ${\cal{S}}$ as defined in Example \ref{pim} (c).
\\
Conversely, for an irreducible discrete copula $D$ with associated  
$L$-dimensional permutation array $A:=(a_{i_{1} \ldots i_{L}})_{i_{1},\ldots,i_{L}=1}^{M}$ of order $M$, we consider the sets  
${\cal{X}}_{1}:=\{x_{1}^{1}< \ldots <x_{M}^{1}\},\ldots,{\cal{X}}_{L}:=\{x_{1}^{L}< \ldots <x_{M}^{L}\}$. Then, $D$ is the empirical copula of the set ${\cal{S}}:=\{(x_{i_{1}}^{1},\ldots,x_{i_{L}}^{L}) | a_{i_{1} \ldots i_{L}}=1\}$.
}
\item[(d)]{The discrete copula $D$ from Example \ref{pim} (d) and Table \ref{tab1}, respectively, corresponds to the three-dimensional stochastic array $A:=(a_{i_1i_2i_3})_{i_1,i_2,i_3=1}^{3}$ of order 3 given in Table \ref{tab2}, which can be verified using Equation \eqref{stochm}. 
\begin{table}[t]
\caption{\small{Entries $a_{i_1i_2i_3}$ of the three-dimensional stochastic array $A:=(a_{i_1i_2i_3})_{i_1,i_2,i_3=1}^{3}$ corresponding to the discrete copula $D$ from Example \ref{pim} (d) and Table \ref{tab1}, respectively.\vspace{0.2 cm}}}\label{tab2}
\centering
\footnotesize
 \begin{minipage}[b]{.325\linewidth}
 \centering
$i_{1}=1$\\
\begin{tabular}{c|ccc}
\toprule
\backslashbox[4mm]{$i_{3}$\kern-1em}{\kern-1em$i_{2}$}&$1$&$2$&$3$\\ \hline 
$1$&$1/4$&$1/4$&$0$\\
$2$&$0$&$0$&$1/4$\\
$3$&$0$&$0$&$1/4$\\
\bottomrule
\end{tabular}
  \end{minipage}	
\begin{minipage}[b]{.325\linewidth}
 \centering
$i_{1}=2$\\
\begin{tabular}{c|ccc}
\toprule
\backslashbox[4mm]{$i_{3}$\kern-1em}{\kern-1em$i_{2}$}&$1$&$2$&$3$\\ \hline 
$1$&$0$&$1/4$&$0$\\
$2$&$1/4$&$0$&$1/4$\\
$3$&$1/4$&$0$&$0$\\
\bottomrule
\end{tabular}
  \end{minipage}	
\begin{minipage}[b]{.325\linewidth}
 \centering
$i_{1}=3$\\
\begin{tabular}{c|ccc}
\toprule
\backslashbox[4mm]{$i_{3}$\kern-1em}{\kern-1em$i_{2}$}&$1$&$2$&$3$\\ \hline 
$1$&$0$&$1/4$&$0$\\
$2$&$1/4$&$0$&$0$\\
$3$&$0$&$1/4$&$1/4$\\
\bottomrule
\end{tabular}
  \end{minipage}
\end{table}
}
\end{enumerate}
\end{example}



\section{A multivariate discrete version of Sklar's theorem}\label{mdcsklar}   

\noindent The key result in the context of copulas undoubtedly is Sklar's theorem \citep{Sklar1959,Nelsen2006}. We now aim at stating and proving a multivariate discrete version thereof.
\\
\\
In the continuous case, an established proof of Sklar's theorem employs an extension lemma, stating that every subcopula can be extended to a copula. The extension lemma in turn is shown via a multivariate interpolation argument \citep[and references therein]{Nelsen2006}. We are guided by this idea and first formulate an extension lemma in a multivariate discrete setting, which provides the main ingredient to showing a multivariate discrete variant of Sklar's theorem. In the proof, which involves rather tedious calculations and is deferred to Appendix \ref{extension.proof}, we employ the one-to-one correspondence of discrete copulas to stochastic arrays from Theorem \ref{equiv}. A bivariate variant of the discrete extension lemma has been shown in \cite{Mayor&2007}.
\begin{lemma} \label{extension} 
(Extension lemma) For each irreducible discrete subcopula $D^{\ast}: J_{M}^{(1)} \times \cdots \times J_{M}^{(L)} \rightarrow I_{M}$, there is an irreducible discrete copula $D: I_{M}^L \rightarrow I_{M}$ such that
\begin{equation*}
D|_{J_{M}^{(1)} \times \cdots \times J_{M}^{(L)}} = D^{\ast},
\end{equation*} 
that is, the restriction of $D$ to $J_{M}^{(1)} \times \cdots \times J_{M}^{(L)}$ coincides with $D^{\ast}$.
\end{lemma}
\noindent Generally, the extension proposed in Lemma \ref{extension} is not uniquely determined.
\\
\\
We are now ready to state a multivariate discrete version of Sklar's theorem, whose proof using Lemma \ref{extension} can be found in Appendix \ref{sklarmdc.proof}. For the bivariate case, such a result can be found in \cite{Mayor&2007}. 
\begin{theorem}\label{sklarmdc} 
(Sklar's theorem in the multivariate discrete case)
\begin{enumerate}
\item{Let $F_{1},\ldots,F_{L}$ be finite univariate CDFs with $\textup{Ran}(F_{\ell}) \subseteq I_{M}$ for all $\ell \in \{1,\ldots,L\}$. If $D$ is an irreducible discrete copula on $I_{M}^{L}$, the function
\begin{equation}\label{sklar}
H(y_{1},\ldots,y_{L}):=D(F_{1}(y_{1}),\ldots,F_{L}(y_{L}))
\end{equation}
for $y_1,\ldots,y_L \in \overline{\mathbb{R}}$ is a finite $L$-dimensional CDF with $\textup{Ran}(H) \subseteq I_{M}$, having $F_{1},\ldots,F_{L}$ as marginal CDFs.
}
\item{Conversely, if $H$ is a finite $L$-dimensional CDF with marginal finite univariate CDFs $F_{1},\ldots,F_{L}$ and  $\textup{Ran}(H) \subseteq I_{M}$, there exists an irreducible discrete copula $D$ on $I_{M}^{L}$ such that
\begin{equation*}
H(y_{1},\ldots,y_{L})=D(F_{1}(y_{1}),\ldots,F_{L}(y_{L}))
\end{equation*}
for $y_1,\ldots,y_L \in \overline{\mathbb{R}}$. Furthermore, $D$ is uniquely determined if $\textup{Ran}(F_{\ell}) = I_{M}$ for all $\ell \in \{1,\ldots,L\}$.
}
\end{enumerate}
\end{theorem}
\noindent Theorem \ref{sklarmdc} is tailored to situations with empirical copulas for data without ties. It is for instance relevant in the context of the applications in Section \ref{applecc}.

\begin{example}\label{die} 
\begin{table}[t]
\caption{\small{A possible discrete copula $D$ for the scenario in Example \ref{die}: The explicit values $D\left(\frac{i_{1}}{6},\frac{i_{2}}{6},\frac{i_{3}}{6}\right)$ are given, where those in bold print are uniquely determined by the corresponding values of the discrete subcopula $D^{\ast}$.\vspace{0.2 cm}}}\label{tab1.exsklar}
\centering
\footnotesize
 \begin{minipage}[b]{.49\linewidth}
 \centering
$i_{1}=0$\\
\begin{tabular}{c|cccc}
\toprule
\backslashbox[4mm]{$i_{3}$\kern-1em}{\kern-1em$i_{2}$}&$0$&$1$&$2$&$3$\\ \hline 
$0$&$\boldsymbol{0}$&$\boldsymbol{0}$&$0$&$\boldsymbol{0}$\\
$1$&$0$&$0$&$0$&$0$\\
$2$&$\boldsymbol{0}$&$\boldsymbol{0}$&$0$&$\boldsymbol{0}$\\
$3$&$\boldsymbol{0}$&$\boldsymbol{0}$&$0$&$\boldsymbol{0}$\\
\bottomrule
\end{tabular}
  \end{minipage}
\begin{minipage}[b]{.49\linewidth} 
\centering
$i_{1}=1$\\
\begin{tabular}{c|cccc}
\toprule
\backslashbox[4mm]{$i_{3}$\kern-1em}{\kern-1em$i_{2}$}&$0$&$1$&$2$&$3$\\ \hline 
$0$&$\boldsymbol{0}$&$\boldsymbol{0}$&$0$&$\boldsymbol{0}$\\
$1$&$0$&$1/3$&$1/3$&$1/3$\\
$2$&$\boldsymbol{0}$&$\boldsymbol{1/3}$&$1/3$&$\boldsymbol{1/3}$\\
$3$&$\boldsymbol{0}$&$\boldsymbol{1/3}$&$1/3$&$\boldsymbol{1/3}$\\
\bottomrule
\end{tabular}
  \end{minipage}
   \begin{minipage}[b]{.49\linewidth} 
   \centering
$i_{1}=2$\\
\begin{tabular}{c|cccc}
\toprule
\backslashbox[4mm]{$i_{3}$\kern-1em}{\kern-1em$i_{2}$}&$0$&$1$&$2$&$3$\\ \hline 
$0$&$\boldsymbol{0}$&$\boldsymbol{0}$&$0$&$\boldsymbol{0}$\\
$1$&$0$&$1/3$&$1/3$&$1/3$\\
$2$&$\boldsymbol{0}$&$\boldsymbol{1/3}$&$1/3$&$\boldsymbol{1/3}$\\
$3$&$\boldsymbol{0}$&$\boldsymbol{1/3}$&$2/3$&$\boldsymbol{2/3}$\\
\bottomrule
\end{tabular}
  \end{minipage}
   \begin{minipage}[b]{.49\linewidth} 
   \centering
   \vspace{0.5 cm}
   $i_{1}=3$\\
\begin{tabular}{c|cccc}
\toprule
\backslashbox[4mm]{$i_{3}$\kern-1em}{\kern-1em$i_{2}$}&$0$&$1$&$2$&$3$\\ \hline 
$0$&$\boldsymbol{0}$&$\boldsymbol{0}$&$0$&$\boldsymbol{0}$\\
$1$&$0$&$1/3$&$1/3$&$1/3$\\
$2$&$\boldsymbol{0}$&$\boldsymbol{1/3}$&$1/3$&$\boldsymbol{2/3}$\\
$3$&$\boldsymbol{0}$&$\boldsymbol{1/3}$&$2/3$&$\boldsymbol{1}$\\
\bottomrule
\end{tabular}
  \end{minipage}
\end{table}
We illustrate the second part of Theorem \ref{sklarmdc} for the case of $L=3$ and $M=3$ and give an example for obtaining a discrete copula associated to a finite three-dimensional CDF with given univariate marginal CDFs. Let $Y_{1}$ be a random variable that is uniformly distributed on the set $\{1,2,3\}$. Moreover, let $Y_{2}$ and $Y_{3}$ be the $\{0,1\}$- and $\{1,2\}$-valued random variables, respectively, defined by
\begin{equation*}
Y_{2} := \begin{cases}
0 & \mbox{if} \,\, Y_{1}\, \, \mbox{takes the value 1,}\\
1 & \mbox{if} \,\, Y_{1} \,\, \mbox{takes the value 2 or 3}
\end{cases}
\end{equation*}
and
\begin{equation*}
Y_{3} :=\begin{cases}
1 & \mbox{if} \,\, Y_{1} \,\, \mbox{takes an odd value,}\\
2 & \mbox{if} \,\, Y_{1} \,\, \mbox{takes an even value}.
\end{cases}
\end{equation*}
The finite three-dimensional CDF $H_{\boldsymbol{Y}}$ for the random vector $\boldsymbol{Y}:=(Y_{1},Y_{2},Y_{3})$ is then given by
\begin{equation*}
H_{\boldsymbol{Y}}(y_{1},y_{2},y_{3}) = \mathbb{P}(Y_{1} \leq y_{1},Y_{2} \leq y_{2},Y_{3} \leq y_{3}),
\end{equation*}
where $y_1,y_2,y_3 \in \overline{\mathbb{R}}$, and $H$ has the marginal CDFs $F_{1}$ of $Y_{1}$ with $\mbox{Ran}(F_{1})=I_{3}$, $F_{2}$ of $Y_{2}$ with $\mbox{Ran}(F_{2})=\left\{0,\frac{1}{3},1\right\}$ and $F_{3}$ of $Y_{3}$ with $\mbox{Ran}(F_{3})=\left\{0,\frac{2}{3},1\right\}$. 
\\
We have $\mbox{Ran}(H)=I_{3}$, and our goal is now to get a discrete copula corresponding to $H$ and $F_1,F_2$ and $F_3$. First, we define the discrete subcopula $D^{\ast}$ via 
\begin{equation*}
D^{\ast}\left(\frac{i_{1}}{3},\frac{i_{2}}{3},\frac{i_{3}}{3}\right):=H(y_{1},y_{2},y_{3}),
\end{equation*}
where $y_1,y_2,y_3 \in \overline{\mathbb{R}}$, are such that $F_{1}(y_{1})=\frac{i_{1}}{3}$, $F_{2}(y_{2})=\frac{i_{2}}{3}$ and $F_{3}(y_{3})=\frac{i_{3}}{3}$. The domain of this discrete subcopula is $J_{3}^{(1)} \times J_{3}^{(2)} \times J_{3}^{(3)}$, where $J_{3}^{(1)}:=I_{3}$, $J_{3}^{(2)}:=\left\{0,\frac{1}{3},1\right\}$ and $J_{3}^{(3)}:=\left\{0,\frac{2}{3},1\right\}$. Due to Lemma \ref{extension}, we can extend the discrete subcopula $D^{\ast}$ defined on $J_{3}^{(1)} \times J_{3}^{(2)} \times J_{3}^{(3)}$ to a discrete copula $D$ defined on $I_{3}^{3}$. However, the discrete extension copula $D$ is not uniquely determined, and there are in general multiple possibilities to complete the missing values of $D$ which are not covered by the values of $D^{\ast}$.
\\
A possible discrete extension copula $D$ for our scenario is shown in Table \ref{tab1.exsklar}, in which we give the explicit values $D\left(\frac{i_{1}}{3},\frac{i_{2}}{3},\frac{i_{3}}{3}\right)$. The values of $D$ which are uniquely determined by the corresponding values of the discrete subcopula $D^{\ast}$ are indicated by the bold font in Table \ref{tab1.exsklar}, whereas the other values are chosen in confirmity with the axioms for discrete copulas.
\end{example}

\section{Applications in probabilistic weather forecasting: Ensemble copula coupling and the Schaake shuffle}\label{applecc}

\noindent Now we relate the concepts and results from the previous sections to multivariate statistical postprocessing techniques for ensemble weather forecasts. Particularly, we study connections to the ensemble copula coupling approach \citep{Schefzik&2013} and the Schaake shuffle \citep{Clark&2004} and thus deepen the considerations in \cite{Schefzik&2013}.
\\
\\
In state-of-the-art meteorological practice, weather forecasts are usually derived from ensemble prediction systems. An ensemble comprises multiple runs of numerical weather prediction models differing in the initial conditions and/or in details of the parameterized numerical representation of the 
atmosphere \citep{GneitingRaftery2005}, thereby addressing the major sources of uncertainty. As ensemble forecasts typically reveal biases and dispersion errors, it is common that they get statistically postprocessed in order to remove these shortcomings. Ensemble predictions and their postprocessing lead to probabilistic forecasts in form of predictive probability distributions 
over future weather quantities. In this context, a forecast distribution of good quality should be as sharp as possible, but on condition of being calibrated \citep{Gneiting&2007}, that is, statistically compatible with the verifying observation. Several 
ensemble postprocessing methods have been proposed, including Bayesian model averaging (BMA) \citep{Raftery&2005} and ensemble model output statistics (EMOS) \citep{Gneiting&2005} as prominent examples. Involved model parameters are typically estimated from a sliding training window consisting of past forecasts and observations. Many of these approaches only 
apply to a single weather
quantity at a single location and  for a single prediction horizon. However, in 
various applications it is crucial to account for spatial, temporal and 
inter-variable dependence
structures.
\\
\\
To address this, several multivariate ensemble postprocessing techniques have been developed. For instance, dependencies between locations can be handled by spatial variants of BMA and EMOS, respectively \citep{Berrocal&2007,Feldmann&2014}. In contrast, the approaches in \cite{Moeller&2013}, \cite{Pinson2012} and \cite{Schuhen&2012} account for purely inter-variable dependencies, with the latter two applying specifically to wind vectors. These methods are parametric and can be interpreted in a Gaussian copula framework \citep{Schefzik&2013}. They are suitable in low-dimensional settings or if a specific structure can be exploited. However, we are often confronted with very high dimensions in weather forecasting. Hence, alternative non-parametric approaches which rely on the use of empirical copulas and are able to address spatial, inter-variable and temporal dependencies simultaneously are of critical interest. Examples for such methods are the ensemble copula coupling (ECC) approach \citep{Schefzik&2013} and the Schaake shuffle \citep{Clark&2004}, which are reviewed and related to the previous results in this paper in what follows, with an emphasis on ECC. 
\\
\\
The ECC method \citep{Schefzik&2013} is based on the rank order information given by the unprocessed raw ensemble forecast. It applies to ensembles consisting members which are exchangeable, that is, statistically indistinguishable, and relies on the mostly plausible assumption that the ensemble is able to represent observed spatial, temporal and inter-variable dependencies appropriately.  To describe ECC formally, let $i \in \{1,\ldots,I\}$ be a weather variable, $j \in \{1,\ldots,J\}$ a location and $k \in \{1,\ldots,K\}$ a prediction horizon, summarized in the multi-index $\ell:=(i,j,k)$. We are given the univariate margins 
\begin{equation}\label{raw.ens}
x_{1}^{\ell},\ldots,x_{M}^{\ell}
\end{equation}
of an $M$-member unprocessed raw ensemble which comprises output in $\mathbb{R}^L$, where the dimension is $L:=I \times J \times K$. Let $\sigma_{\ell}(m):=\operatorname{rank}(x_{m}^{\ell})$, where $m \in \{1,\ldots,M\}$, be the permutation of $\{1,\ldots,M\}$ induced by the order statistics $x_{(1)}^{\ell} \leq \ldots \leq x_{(M)}^{\ell}$ of the raw ensemble for each fixed margin $\ell$, with any ties resolved at random. Our goal is to come up with the postprocessed ECC ensemble $\hat{x}_{1}^{\ell},\ldots,\hat{x}_{M}^{\ell}$ of the same size $M$ as the raw ensemble.
\\
To achieve this, we first apply a state-of-the-art univariate postprocessing method, such as BMA or EMOS, to the raw ensemble $x_{1}^{\ell},\ldots,x_{M}^{\ell}$ and obtain a postprocessed predictive CDF $F_{\ell}$ for each variable, location and look-ahead time individually. Then, each CDF $F_\ell$ is represented by a discrete sample 
\begin{equation}\label{discr.sample}
\tilde{x}_{1}^{\ell},\ldots,\tilde{x}_{M}^{\ell}
\end{equation}
of size $M$. For instance, this can be conveniently done by taking the equally spaced $\frac{1}{M+1}-,\ldots,\frac{M}{M+1}-$\\quantiles of $F_{\ell}$. 
\\
In the final ECC step, the sample $\tilde{x}_{1}^{\ell},\ldots,\tilde{x}_{M}^{\ell}$ is rearranged for each margin $\ell$ with respect to the ranks of the raw ensemble members. That is, the final postprocessed ECC ensemble $\hat{x}_{1}^{\ell},\ldots,\hat{x}_{M}^{\ell}$ is for each $\ell$ given by
\begin{equation}\label{ecc.step}
\hat{x}_{1}^{\ell}:=\tilde{x}_{(\sigma_{\ell}(1))}^{\ell},\ldots,\hat{x}_{M}^{\ell}:=\tilde{x}_{(\sigma_{\ell}(M))}^{\ell}.
\end{equation} 
The last reordering step \eqref{ecc.step} is essential, as it transfers the spatial, inter-variable and temporal rank dependence structure of the raw ensemble to the postprocessed ECC ensemble, thereby capturing the flow dependence. If we simply stopped the procedure at the sampling stage \eqref{discr.sample}, we would obtain a postprocessed ensemble which does not conserve dependence patterns. Such an ensemble will be referred to as an individually postprocessed ensemble in what follows.
\\
ECC performed well in several case studies \citep{Schefzik&2013,Feldmann&2014,Schefzik2014}. Since its crucial reordering step is computationally non-expensive, one of the major advantages of ECC is that it practically comes for free, once the univariate postprocessing has been performed. However, BMA and EMOS as univariate postprocessing methods are already implemented in the R packages \texttt{ensembleBMA} and \texttt{ensembleMOS}, respectively, which are freely available at \url{http://cran.r-project.org} \citep{R2011}. In addition, ECC offers a simple and intuitive, yet powerful technique that goes without complex modeling or sophisticated parameter fitting, thus providing a natural benchmark. It combines analytical, statistical and numerical modeling and is generally applicable to much broader settings apart from weather forecasting \citep{Schefzik&2013}. 
\\
\\
As indicated by its name, ECC has strong connections to copulas, particularly to the notions and results presented before. To deepen this, let  
$X_{1},\ldots,X_{L}$ be discrete random variables taking values in
$\{x_{1}^{1},\ldots,x_{M}^{1}\},\ldots,\{x_{1}^{L},\ldots,x_{M}^{L}\}$,  
respectively, where $x_{1}^{\ell},\ldots,x_{M}^{\ell}$ for $\ell \in \{1,\ldots,L\}$ is the $M$-member  
raw ensemble forecast from above. For convenience, we assume that there are no ties among the  
corresponding raw ensemble margins. Concerning the multivariate random vector  
$\boldsymbol{X}:=(X_{1},\ldots,X_{L})$, the respective univariate CDFs  
$R_{1},\ldots,R_{L}$ take values in $I_{M}$, that is,  
$\mbox{Ran}(R_{{1}})=\cdots=\mbox{Ran}(R_{{L}})=I_{M}$. For the multivariate CDF $R: \mathbb{R}^L \rightarrow I_M$ of $\boldsymbol{X}$, we also have $\mbox{Ran}(R)=I_{M}$. According to the multivariate discrete  
version of Sklar's theorem in Theorem \ref{sklarmdc} tailored to such a framework, there exists a uniquely  
determined irreducible discrete, and hence a uniquely determined empirical, copula $E_M:I_{M}^{L} \rightarrow I_{M}$
such that
\begin{equation}\label{raw.interpr}
R(y_{1},\ldots,y_{L}) = E_M(R_{1}(y_{1}),\ldots,R_{L}(y_{L}))
\end{equation}
for $y_{1},\ldots,y_{L} \in \overline{\mathbb{R}}$. That is, the multivariate distribution $R$ is connected to its univariate
margins $R_1,\ldots,R_L$ via $E_M$. Conversely, if we take $E_M$ to be the empirical copula induced by  
the raw ensemble forecast  
$\{x_1^1,\ldots,x_M^1\},\ldots,\{x_1^L,\ldots,x_M^L\}$ and  
$R_1,\ldots,R_L$ to be the univariate CDFs of the raw ensemble  
margins, then $R$ as constructed in \eqref{raw.interpr} is a  
multivariate CDF.
\\
\\
Following and generalizing the statistical interpretation of discrete  
copulas for the bivariate case in \cite{Mesiar2005},
\begin{equation*}
E_M\left(\frac{i_{1}}{M},\ldots,\frac{i_{L}}{M}\right)=\mathbb{P}(R \in  
[-\infty,y_{1}]  \times \cdots \times [-\infty,y_{L}]),
\end{equation*}
where $y_{1},\ldots,y_{L} \in \overline{\mathbb{R}}$ such that 
\begin{equation*}
R_{1}(y_{1})=\mathbb{P}(X_{1} \leq y_{1})=\frac{i_{1}}{M},\ldots,R_{L}(y_{L})=\mathbb{P}(X_{L} \leq y_{L})=\frac{i_{L}}{M},
\end{equation*}
that is, 
\begin{equation*}
E_M\left(\frac{i_{1}}{M},\ldots,\frac{i_{L}}{M}\right)=\mathbb{P}(X_{1} \leq y_{1},\ldots,X_{L} \leq y_{L}).
\end{equation*}
To describe the discrete probability distribution of the random vector $\boldsymbol{X}$, we set 
\begin{equation*}
\alpha_{i_{1} \ldots i_{L}} := \mathbb{P}(X_{1}=x_{(i_{1})}^{1},\ldots,X_{L}=x_{(i_{L})}^{L}),
\end{equation*}
where $x_{(i_{\ell})}^{\ell}$ for $i_{\ell} \in \{1,\ldots,M\}$ and $\ell \in \{1,\ldots,L\}$ denote the corresponding order statistics of the values $X_{1},\ldots,X_{L}$ attain. Then, $\alpha_{i_{1} \ldots i_{L}} \in \left\{0,\frac{1}{M}\right\}$ for all $i_{1},\ldots,i_{L} \in \{1,\ldots,M\}$. Hence, $a_{i_{1} \ldots i_{L}}:=M \alpha_{i_{1} \ldots i_{L}} \in \{0,1\}$ for $i_{1},\ldots,i_{L} \in \{1,\ldots,M\}$, $A:=(a_{i_{1} \ldots i_{L}})_{i_{1},\ldots,i_{L}=1}^{M}$ is a permutation array of order $M$, and
\begin{equation*}
\frac{1}{M} \sum\limits_{\nu_{1}=1}^{i_{1}} \cdots \sum\limits_{\nu_{L}=1}^{i_{L}} a_{\nu_{1} \ldots \nu_{L}} = E_M \left(\frac{i_{1}}{M},\ldots,\frac{i_{L}}{M}\right),
\end{equation*}
in accordance with Theorem \ref{equiv}.
\\
\\
In our setting, the above  
considerations hold analogously for both an individually postprocessed  
ensemble $\tilde{x}_1^\ell,\ldots,\tilde{x}_M^\ell$ as in \eqref{discr.sample} and the  
ECC ensemble $\hat{x}_1^\ell,\ldots,\hat{x}_M^\ell$ as in \eqref{ecc.step}. In obvious  
notation, let $\tilde{F}$ and $\hat{F}$ be the respective multivariate  
empirical CDFs. Moreover, let $\tilde{F}_1,\ldots,\tilde{F}_L$ denote  
the marginal empirical CDFs of the individually postprocessed  
ensemble, with $\tilde{E}_M$ being the corresponding empirical copula. Then,
\begin{equation}\label{indivpost.interpr}
\tilde{F}(y_1,\ldots,y_L)=\tilde{E}_M(\tilde{F}_1(y_1),\ldots,\tilde{F}_L(y_L))
\end{equation}
and
\begin{equation}\label{ecc.interpr}
\hat{F}(y_1,\ldots,y_L)=E_M(\tilde{F}_1(y_1),\ldots,\tilde{F}_L(y_L))
\end{equation}
for $y_1,\ldots,y_L \in \overline{\mathbb{R}}$.
\\
\\
Comparing Equations \eqref{raw.interpr}, \eqref{indivpost.interpr} and  
\eqref{ecc.interpr}, the individually postprocessed ensemble and the  
ECC ensemble have the same marginal distributions, whereas the raw  
ensemble and the ECC ensemble are associated with the same empirical  
copula modeling the dependence, due to the design of ECC aiming at  
retaining the rank dependence pattern from the raw ensemble. In particular, the ECC ensemble conserves the bivariate Spearman rank correlation coefficients in the raw ensemble output.
\begin{figure}[p]
\centering
\subfiguretopcaptrue
\subfigure[ECMWF Raw Ensemble]{\includegraphics[scale=0.305]{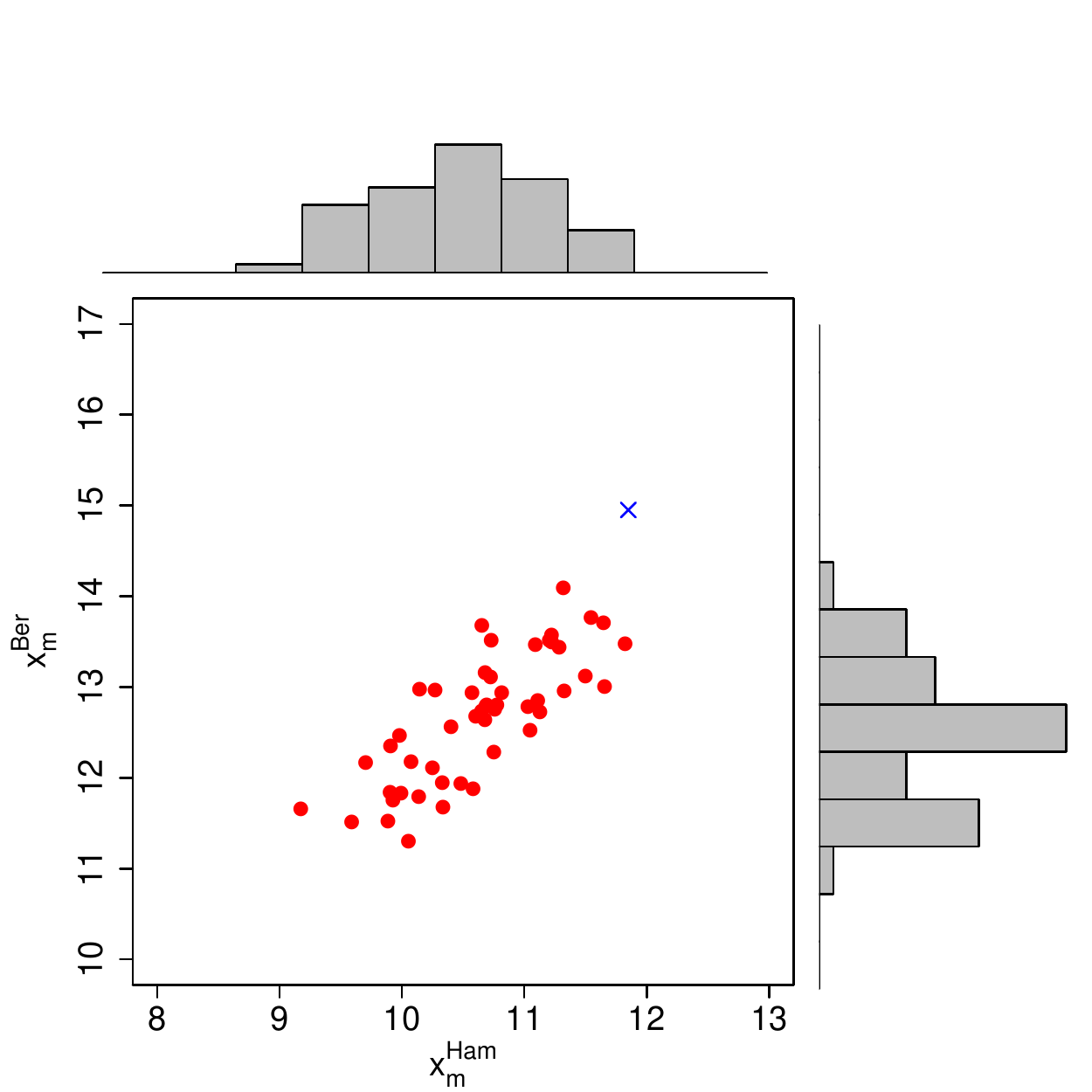}}\hspace{1 cm}
\subfigure[Individually BMA Postprocessed Ensemble]{\includegraphics[scale=0.305]{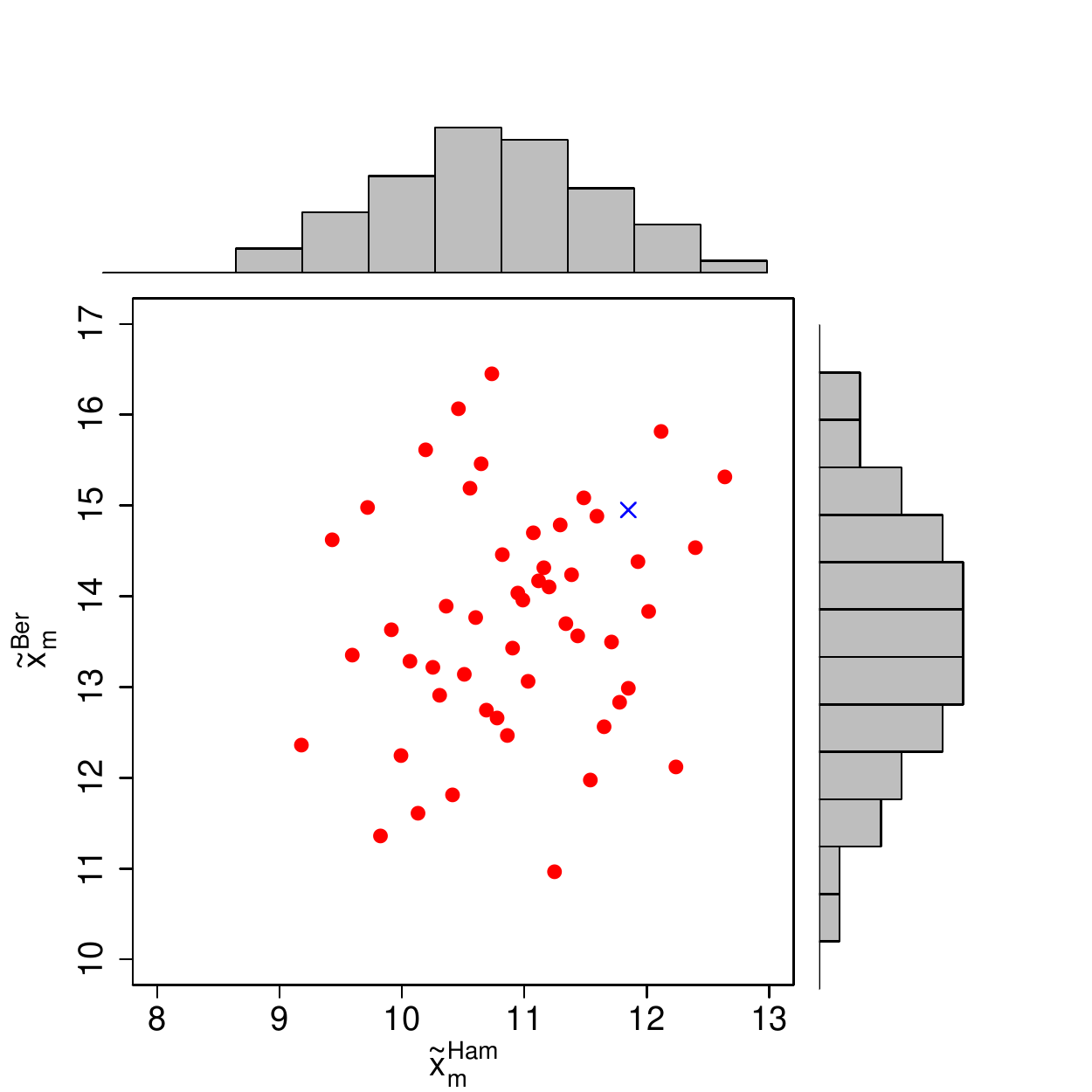}}\hspace{1 cm}
\subfigure[ECC Ensemble]{\includegraphics[scale=0.305]{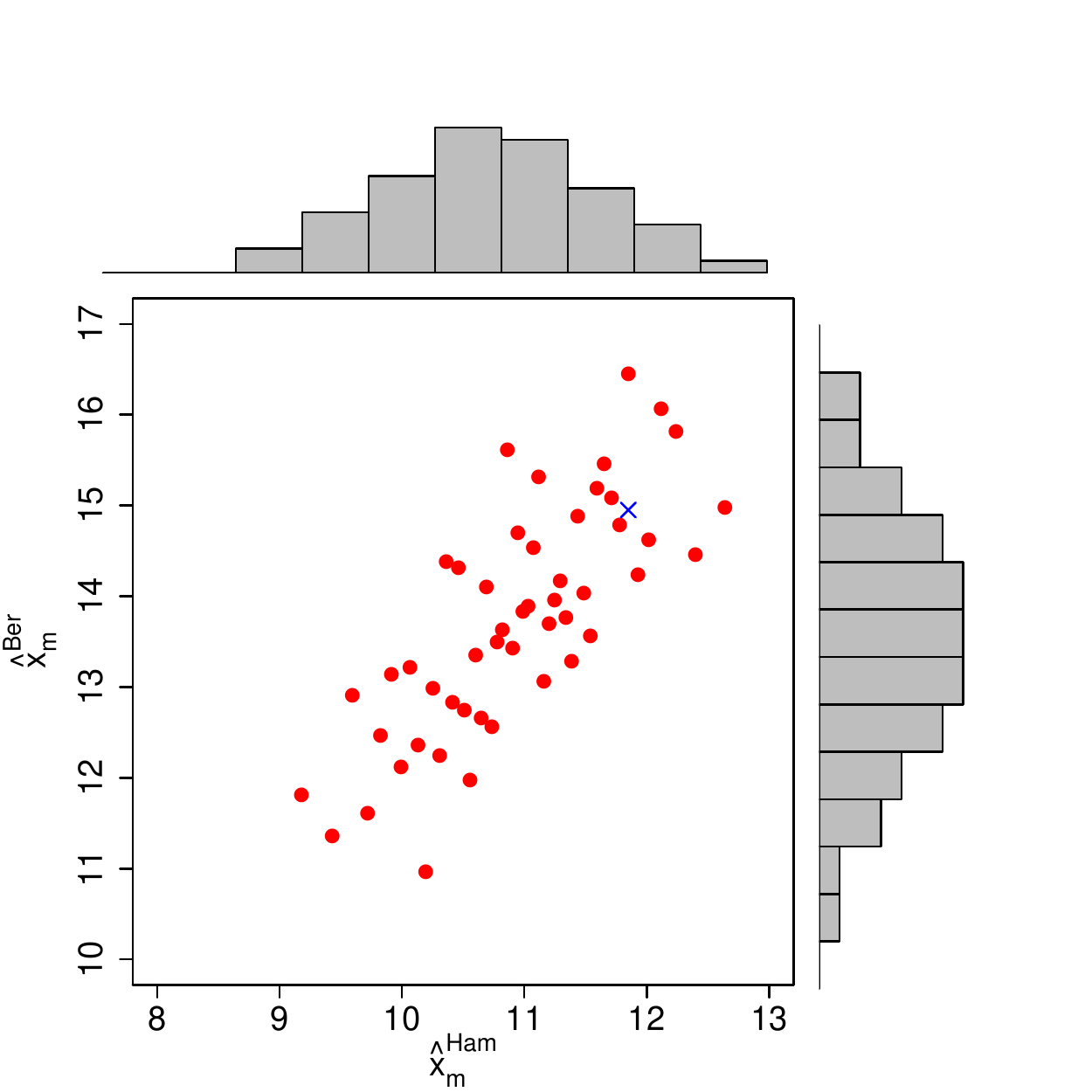}}\newline
\subfigure{\includegraphics[scale=0.22]{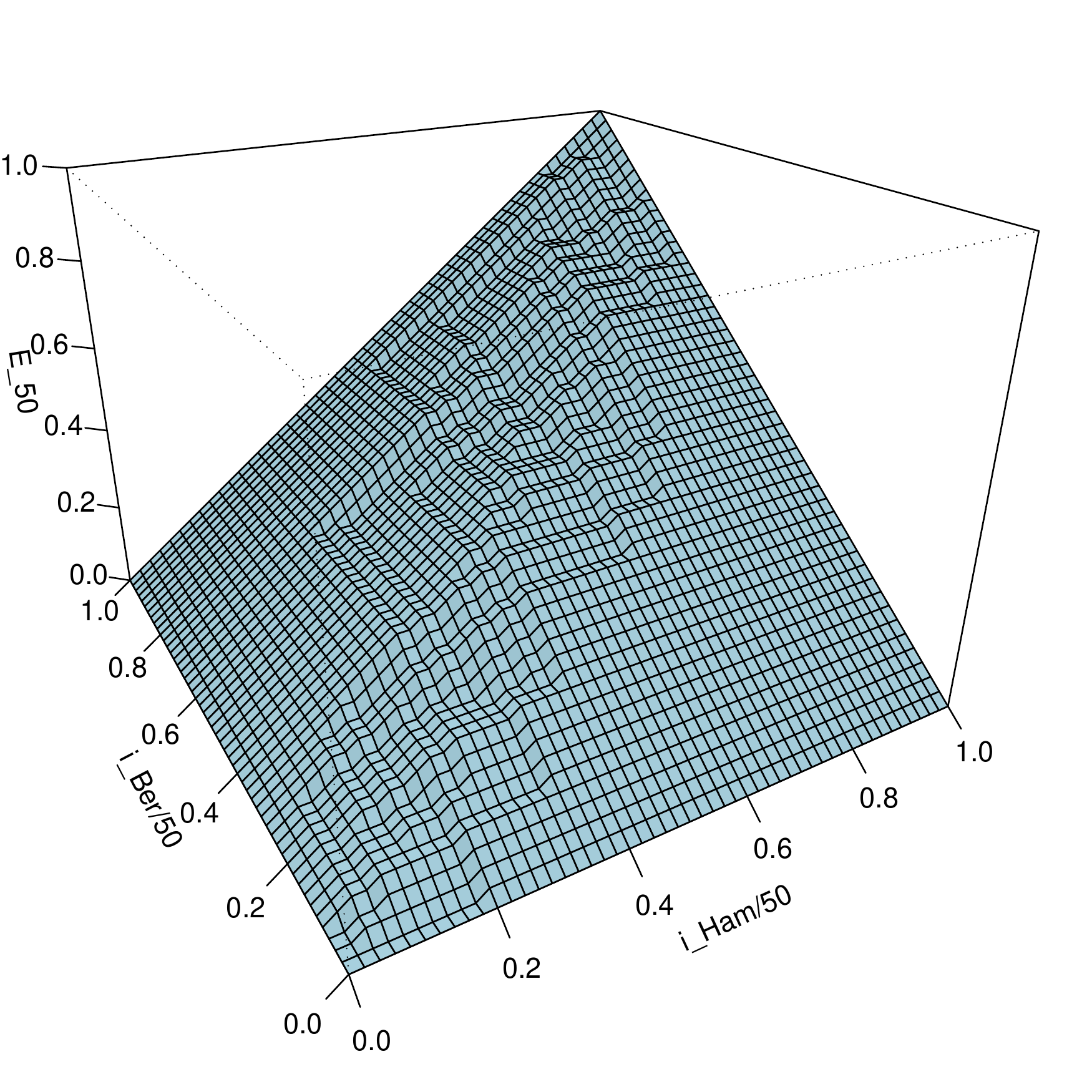}}\hspace{1 cm}
\subfigure{\includegraphics[scale=0.22]{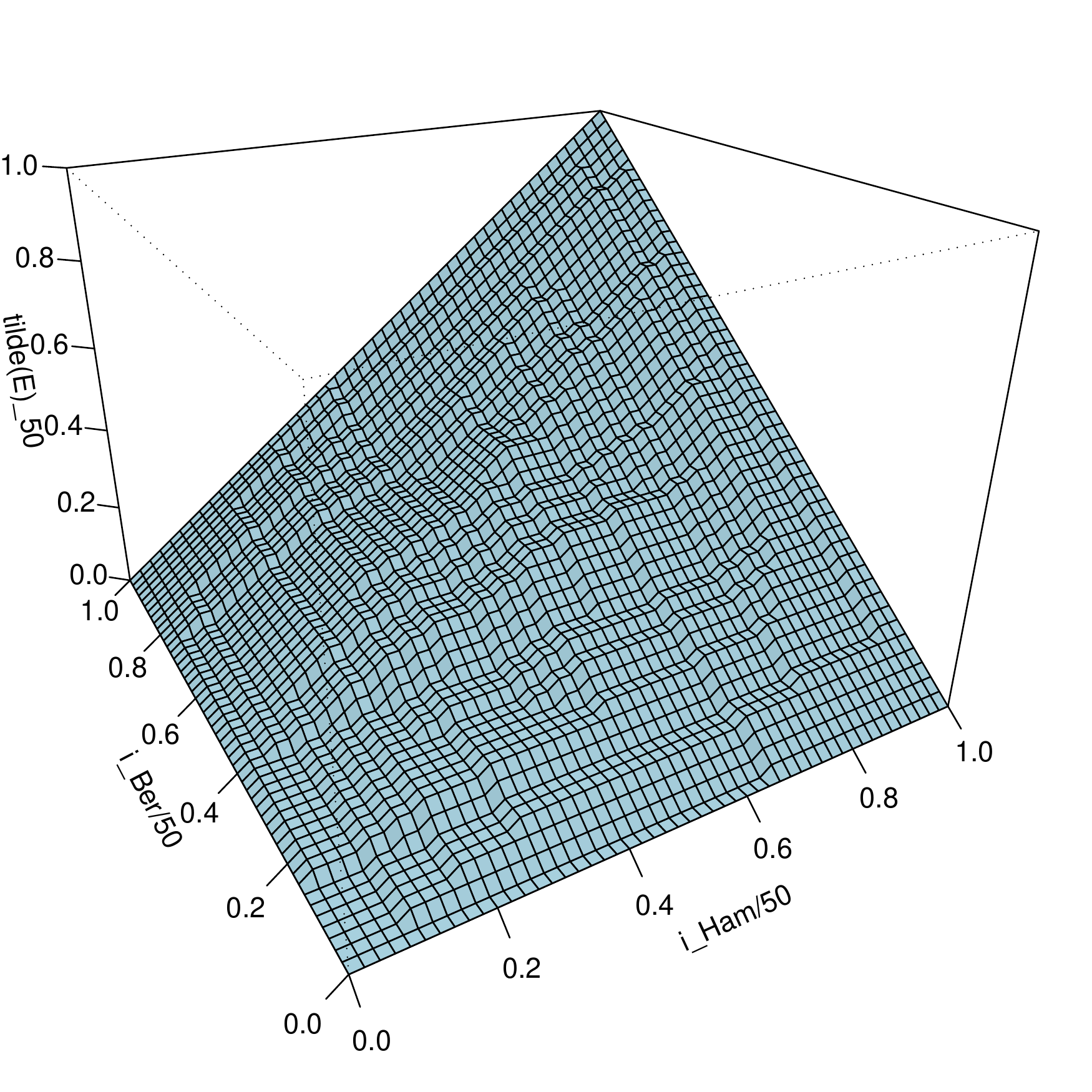}}\hspace{1 cm}
\subfigure{\includegraphics[scale=0.22]{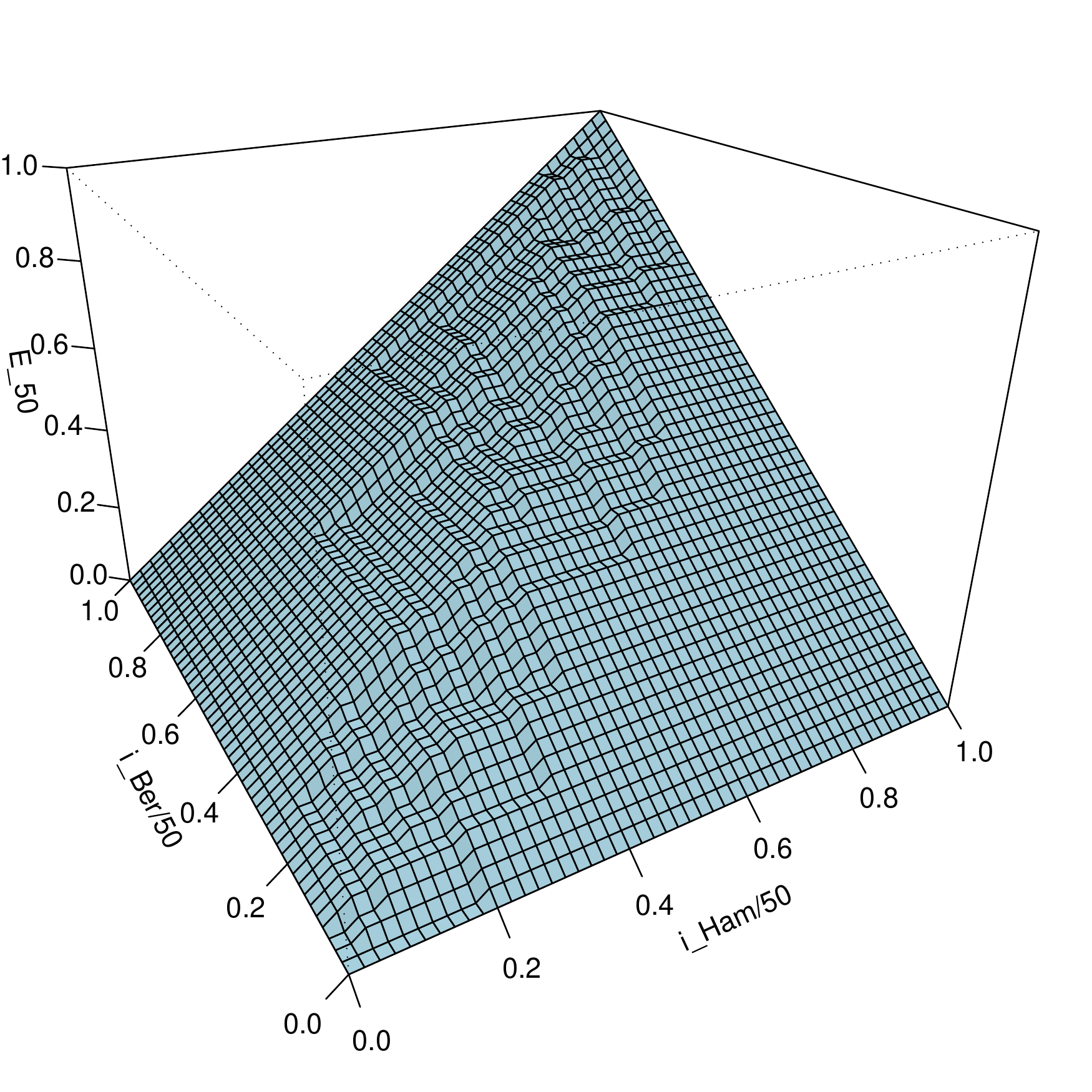}}\newline
\subfigure{\includegraphics[scale=0.3]{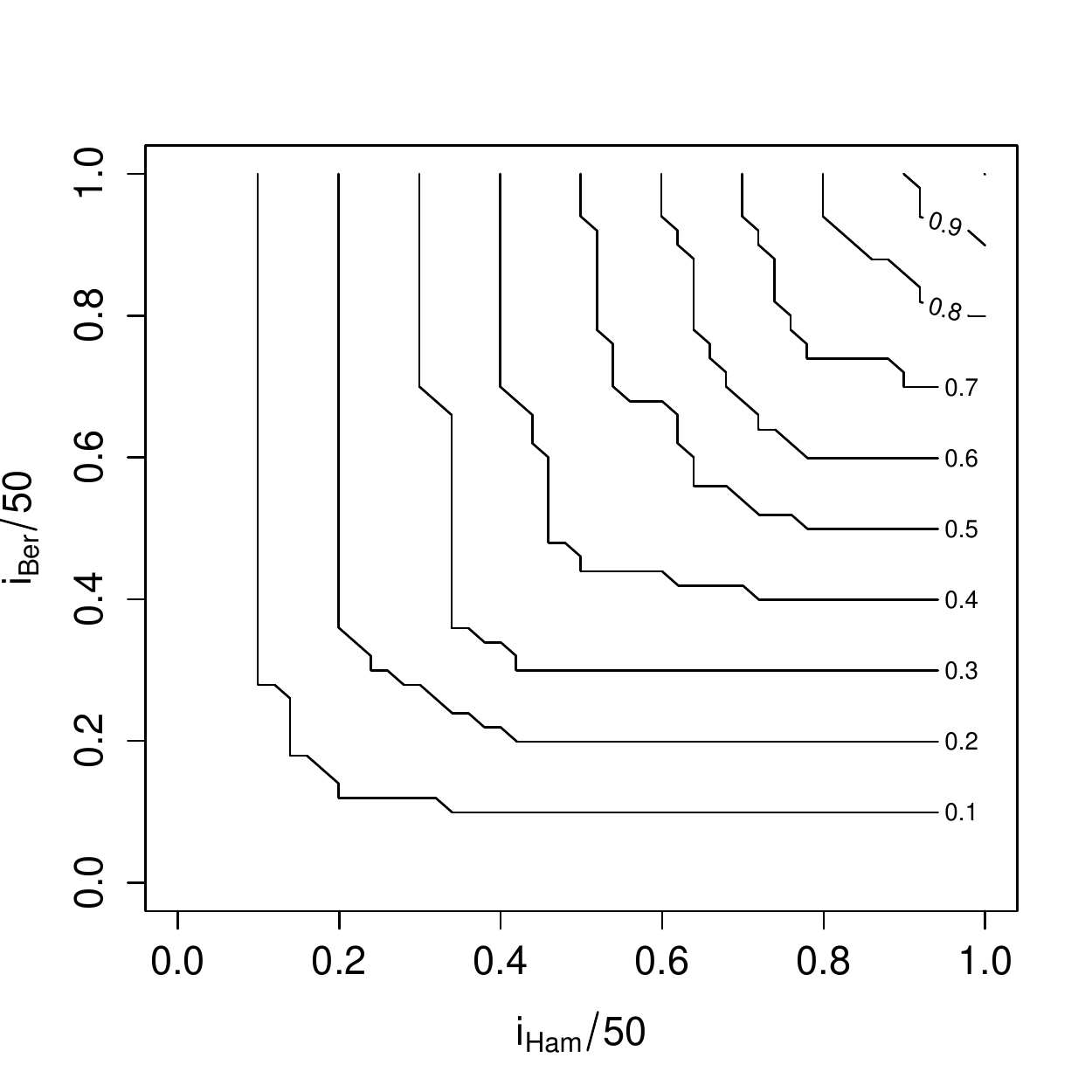}}\hspace{1 cm}
\subfigure{\includegraphics[scale=0.3]{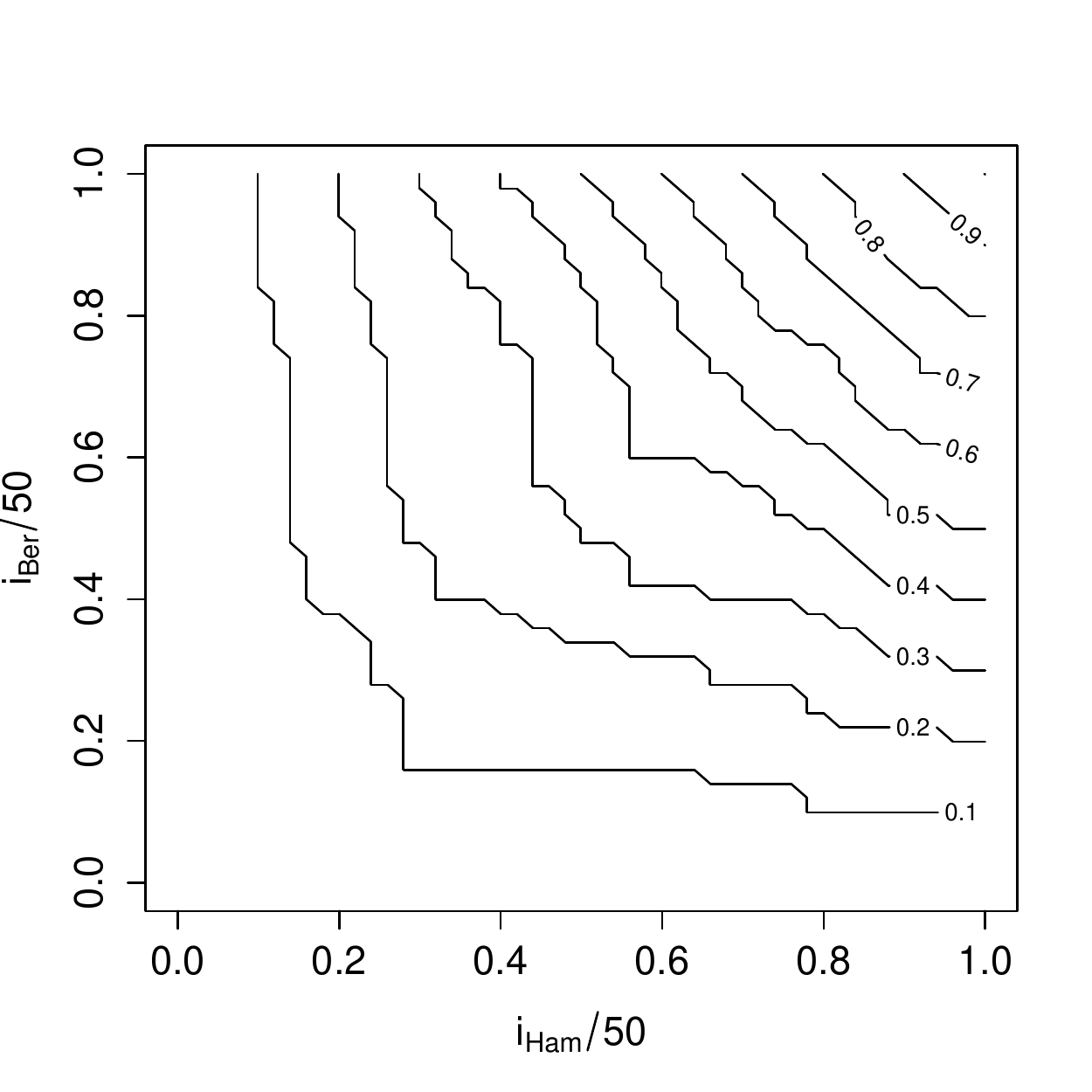}}\hspace{1 cm}
\subfigure{\includegraphics[scale=0.3]{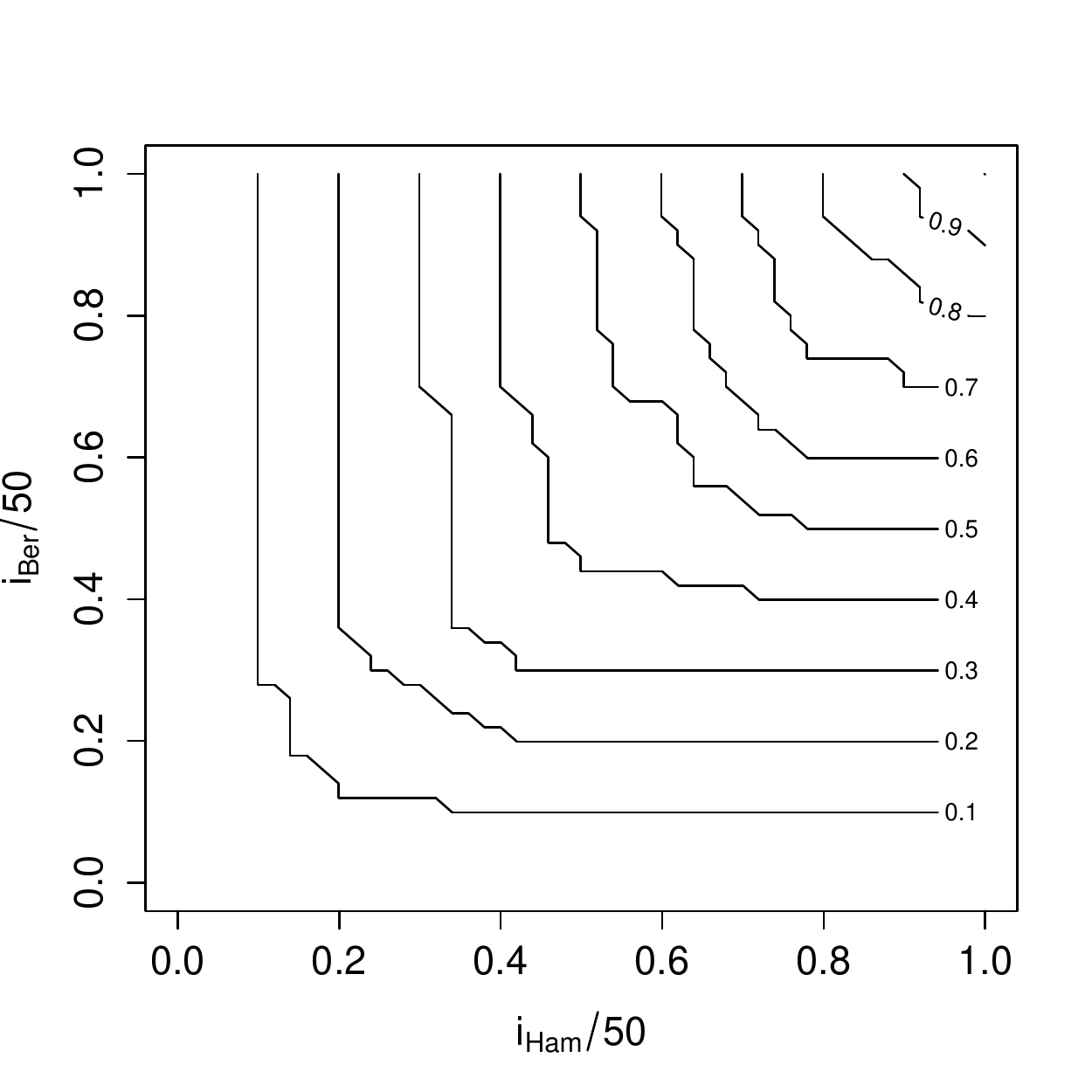}}\newline
\subfigure{\includegraphics[scale=0.3]{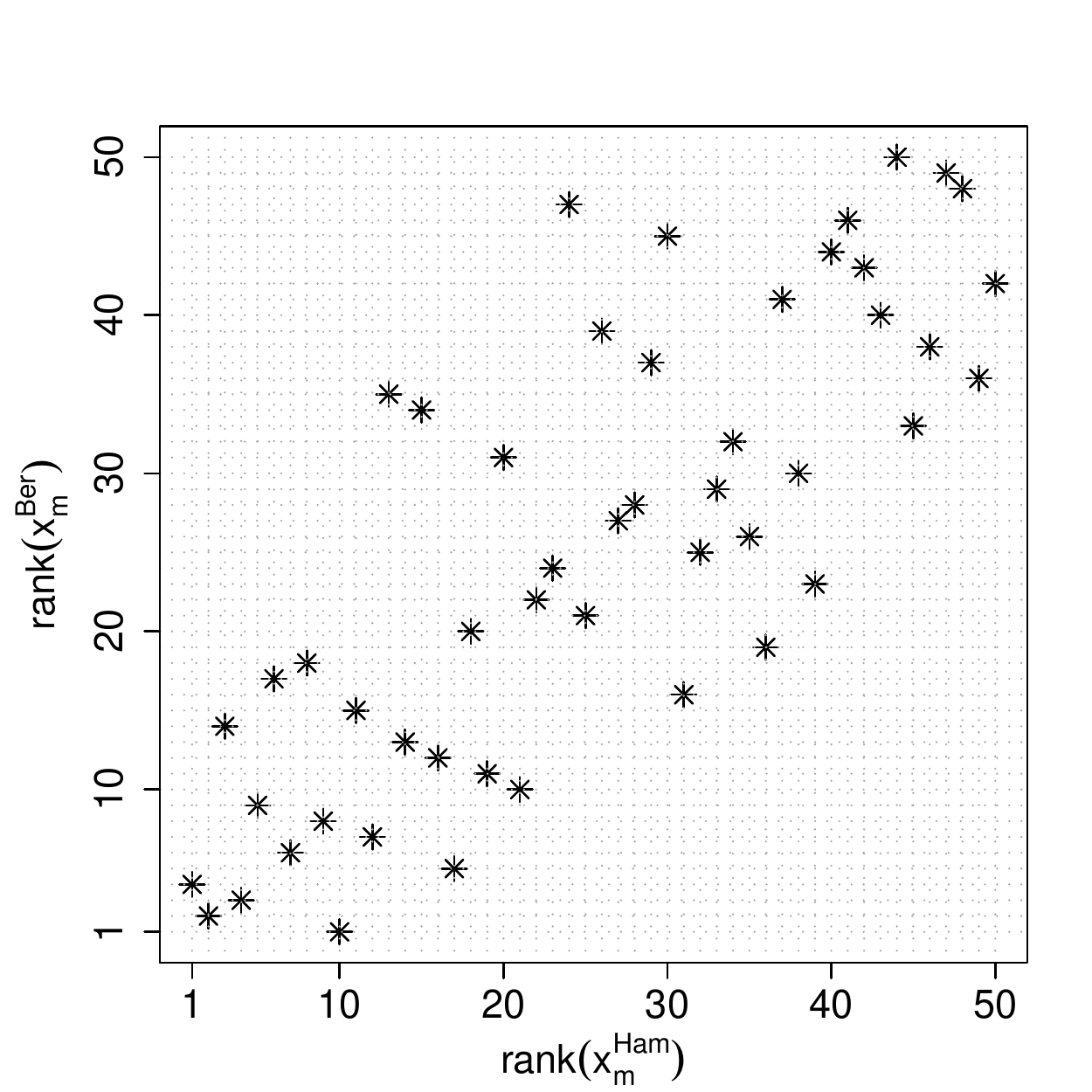}}\hspace{1 cm}
\subfigure{\includegraphics[scale=0.3]{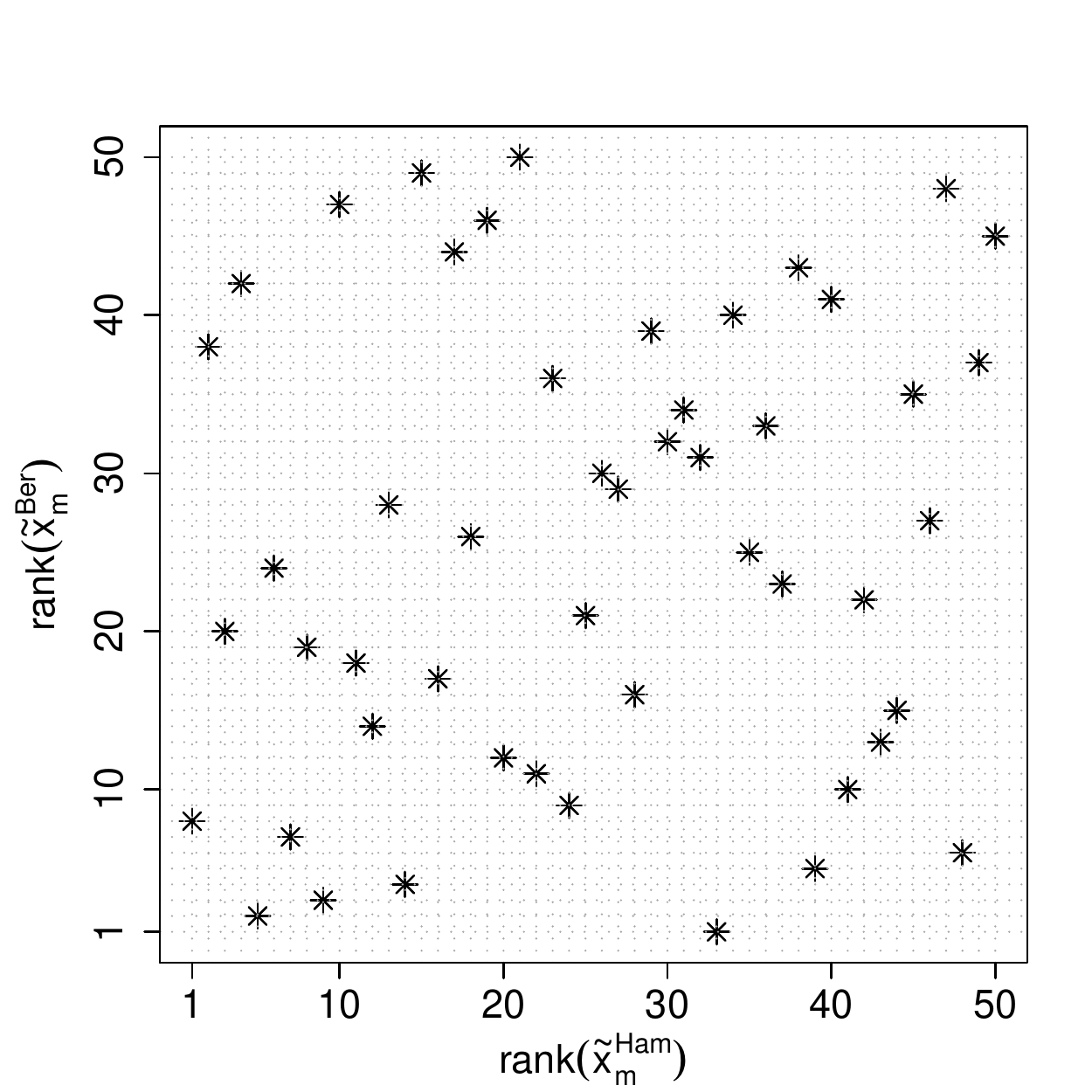}}\hspace{1 cm}
\subfigure{\includegraphics[scale=0.3]{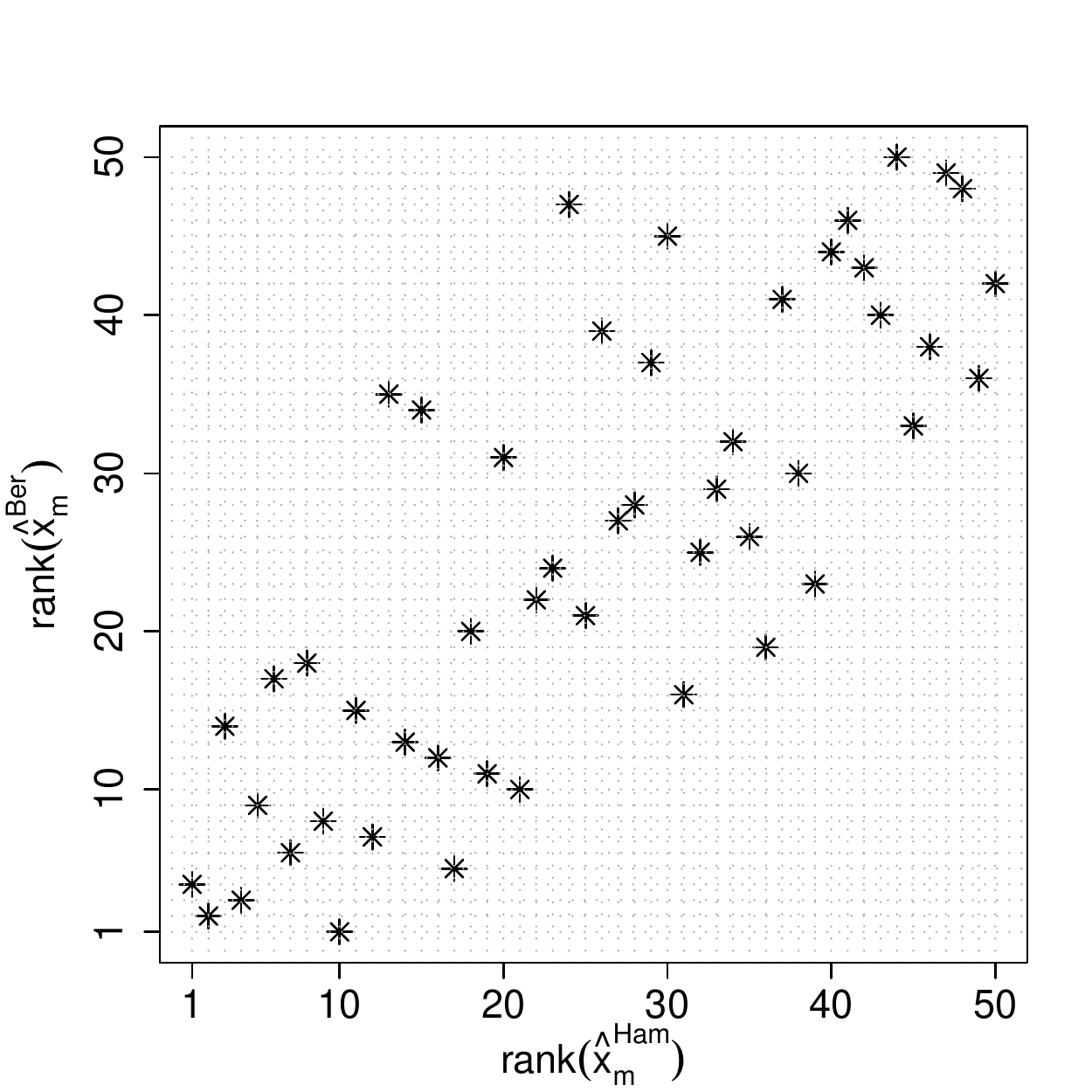}}\newline
\caption{24 hour ahead ensemble predictions for temperature (in $^\circ$C) at Berlin (Ber) and Hamburg (Ham), valid 2:00 am on 27 June 2010, comprising (a) the ECMWF raw, (b) an individually BMA postprocessed and (c) the ECC ensemble. First row: Scatterplots with marginal histograms; red dots: corresponding ensemble forecast, blue cross: verifying observation. Second row: Perspective plots of the corresponding empirical copulas. Third row: Contour plots of the corresponding empirical copulas. Fourth row: Corresponding Latin squares.}
\label{ecc}
\end{figure}
\\
\\
To illustrate the ECC approach and its relations to multivariate discrete copulas, we consider a real data example with forecasts provided by the $M=50$-member European Centre for Medium-Range Weather Forecasts (ECMWF) ensemble \citep{ECMWF2012}. Although our focus has been on the general multivariate case in this paper, we consider for illustrative purposes a bivariate setting, that is, $L=2$, dealing with 24 hour ahead forecasts for temperature (in degrees Celsius; $^\circ$C) at Berlin and Hamburg, valid 2:00 am on 27 June 2010. In the left panel of the first row of Figure \ref{ecc}, the unprocessed raw ensemble forecast according to \eqref{raw.ens} is shown, where each red dot represents an ensemble member $m \in \{1,\ldots,50\}$, and the verifying observation is indicated by the blue cross. The ensemble members reveal a pronounced positive correlation. The mid-plot in the first row shows an individually postprocessed ensemble as in \eqref{discr.sample}, where the univariate postprocessing has been performed via BMA \citep{Raftery&2005} here. While correcting for biases and dispersion errors, the individually site-by-site BMA postprocessed ensemble essentially provides no correlation structure, in that the bivariate rank order characteristics of the unprocessed forecast from the left panel are lost. Finally, the postprocessed ECC ensemble according to \eqref{ecc.step} in the right panel of the first row corrects for biases and dispersion errors as the individually BMA postprocessed ensemble does, but additionally conserves the rank dependence pattern given by the raw ensemble. Thus, although the individually BMA postprocessed and the ECC ensemble have the same marginal distributions, as indicated by the histograms, they differ drastically in their multivariate rank dependence structures. 
\\
In the second and third row of Figure \ref{ecc}, the perspective and stabilized contour plots, respectively, of the empirical copulas linked to the different ensembles in our illustrative example are shown. As discussed before, the raw and the ECC ensemble are associated with the same empirical copula $E_{50}$. On the other hand, the individually BMA postprocessed ensemble is linked to a distinct empirical copula $\tilde{E}_{50}$, whose plots meaningfully resemble those of the independence copula $\Pi$ from Example \ref{pim} (a). According to the equivalences discussed in Section \ref{dcstar}, the raw and the ECC ensemble are also related to the same Latin square of order $M=50$, which is a Latin hypercube \citep{Gupta1974} in $L=2$ dimensions, whereas the individually BMA postprocessed ensemble is not, as illustrated in the fourth row in Figure \ref{ecc}.
\\
\\
In a nutshell, ECC indeed can be considered a discrete copula approach. It comes up with a postprocessed, discrete $L$-dimensional distribution, which is according to Equation \eqref{ecc.interpr} constructed from univariate empirical CDFs $\tilde{F}_{1},\ldots,\tilde{F}_{L}$ and an empirical copula $E_M$. While the marginal CDFs $\tilde{F}_{1},\ldots,\tilde{F}_{L}$ are defined by the samples drawn from the predictive CDFs $F_1,\ldots,F_L$ obtained by univariate postprocessing, the empirical copula $E_M$, which models the dependence structure, is induced by the unprocessed raw ensemble. 
\\
\\
In close analogy to ECC and with very similar justifications, the Schaake shuffle \citep{Clark&2004}, which is an established concept in meteorology, can also be interpreted as a discrete copula approach \citep{Schefzik&2013}. However, the main difference is that the corresponding empirical copula in the Schaake shuffle method is defined based on historical verifying observations, rather than on a raw ensemble forecast as in the ECC approach. Moreover, the size $N$ of the postprocessed Schaake shuffle ensemble is in contrast to that of the ECC ensemble not restricted to equal the size $M$ of the raw ensemble.
\\
To describe the Schaake shuffle formally, let $\ell^\ast:=(i,j)$ denote a multi-index comprising a fixed weather variable $i \in \{1,\ldots,I\}$ at a fixed location $j \in \{1,\ldots,J\}$, and let $L^\ast:=I \times J$. Suppose further that for each margin $\ell^\ast \in \{1,\ldots,L^\ast\}$, we have a univariate postprocessed predictive CDF $F_{\ell^\ast}$, which is represented by a discrete sample $\tilde{x}_1^{\ell^\ast},\ldots,\tilde{x}_N^{\ell^\ast}$ of size $N$. Moreover, let $o_1^{\ell^\ast},\ldots,o_N^{\ell^\ast}$ denote a set of $N$ historical weather observations, where the same $N$ verification dates have to be taken for all margins $\ell^\ast$. These past observations define an empirical copula $O_N$. The multivariate empirical CDF $\hat{F}$ of the postprocessed Schaake shuffle ensemble is then given by analogy with Equation \eqref{ecc.interpr} by
\begin{equation*}
\hat{F}(y_1,\ldots,y_{L^\ast})=O_N(\tilde{F}_1(y_1),\ldots,\tilde{F}_{L^\ast}(y_{L^\ast}))
\end{equation*}
for $y_1,\ldots,y_{L^\ast} \in \overline{\mathbb{R}}$, with the empirical marginal CDFs $\tilde{F}_1,\ldots,\tilde{F}_{L^\ast}$ based on the samples $\tilde{x}_1^{\ell^\ast},\ldots,\tilde{x}_N^{\ell^\ast}$ for $\ell^\ast \in \{1,\ldots,L^\ast\}$. That is, the empirical copula of the historical weather field record is applied to the discrete samples from the univariate postprocessed predictive CDFs. The thus reordered forecast in the Schaake shuffle ensemble consequently inherits the multivariate rank dependence pattern as well as the pairwise Spearman rank correlation coefficients from the underlying historical weather record, rather than from the raw ensemble as in the ECC approach.
\\
\\
In practice, both ECC and the Schaake shuffle have performed well in case studies \citep{Clark&2004,Schefzik&2013,Schefzik2014,ScheuererHamill2015,VracFriederichs2014}, and consequently, both are valuable tools for multivariate ensemble postprocessing that can be used as a benchmark. However, the standard ECC implementation as presented in this paper only applies to raw ensembles consisting of exchangeable members. Moreover, the size of the postprocessed ECC ensemble is restricted to equal that of the raw ensemble. When employing ECC, the raw ensemble should thus be reasonably large to have reliable information about the dependence structure. In contrast, the Schaake shuffle can principally create a postprocessed ensemble of arbitrary size, provided that there is a sufficiently large database of observations. However, its standard implementation fails to address atmospheric flow and time dependence, unlike ECC.
\\
The development of modified variants of ECC and the Schaake shuffle that are able to solve the shortcomings mentioned above is an issue of current research, where first attempts can be found in \cite{Schefzik2014} and \cite{Wilks2014}. 
\\
The question whether to prefer ECC or the Schaake shuffle cannot be definitely answered yet, as investigations on this are still in progress and not exhausted. A direct comparison of the predictive performances of the two methods has hitherto been conducted in a recent paper by \citet{Wilks2014} only. In many of the settings considered in \cite{Wilks2014}, but not in all, the Schaake shuffle outperforms ECC. However, the case study in \cite{Wilks2014} is based on a comparably small raw ensemble of size $M=11$, and ECC would be expected to perform better and similarly well as the Schaake shuffle if the raw ensemble was larger. A reimplementation of Wilks' initial case study \citep{Wilks2014} using an unprocessed ensemble consisting of more members, such as the $M=50$-member ECMWF ensemble, together with a comparison to the ECC modificatins in \cite{Schefzik2014}, is therefore of great interest.

\section{Discussion}\label{discussion}

In this paper, we have extended the notion of discrete copulas \citep{Mayor&2005,Kolesarova&2006} from the bivariate to the multivariate case. Moreover, we have shown the equivalence of multivariate discrete copulas to stochastic arrays and have proven a multivariate discrete version of Sklar's theorem.
\\
\\
The theoretical frame, in which the presented results have been developed, is driven by and tailored to the applications in meteorology considered in the last section. In particular, we focused on data with no ties, and multivariate discrete copulas have been defined on the domain $I_M^L$ and thus on points that are equally spaced across $I_M^L$. Hence, the discrete copula concept could be generalized to settings in which either the domain is $I_{M_1} \times \cdots \times I_{M_L}$ , where $M_1,\ldots,M_L \in \mathbb{N}$ are distinct, or the points where the copula is of interest are heterogeneously spaced across the margins.
\\
\\
The second part of the multivariate discrete version of Sklar's theorem as formulated in Theorem \ref{sklarmdc} can very likely be extended and accentuated to the effect that the irreducible discrete copula $D$ is uniquely determined if and only if $\operatorname{Ran}(F_\ell)=I_M$ for all $\ell \in \{1,\ldots,L\}$. Following the procedure for the bivariate case in \cite{Mayor&2007}, a rigorous proof of this would require the construction of a smallest and largest discrete extension copula, respectively. The corresponding algorithms to achieve this appear to be supplied by modifying those for the bivariate case in \cite{Mayor&2007}. However, due to its expected length and technical character, a detailed study of this question has not been conducted here and is therefore an issue for future work.
\\
\\
Although several multivariate copula-based methods for discrete data have already been proposed, for instance in \cite{Panagiotelis&2012} using vine and pair copulas, approaches based on multivariate discrete copulas as introduced in this paper provide appropriate and useful alternatives. As we have seen, the notion of discrete copulas quite naturally arises in the context of the ECC method and the Schaake shuffle related to statistical ensemble postprocessing in weather forecasting. It allows for a unifying interpretation and builds an overarching theoretical frame of many other state-of-the-art ensemble postprocessing techniques which can be considered special cases of ECC \citep{Schefzik&2013}. Examples include the methods in \cite{Pinson2012}, \cite{RoulinVannitsem2012} and \cite{VanSchaeybroeckVannitsem2013}, for instance. The ECC scheme principally can also be employed in applications apart from weather prediction, and the theoretical frame then continues to hold analogously in much broader situations.
\\
\\
Following the pioneering work of \citet{Wilks2014}, further research should be conducted towards a comparison of the predictive performances of ECC, the Schaake shuffle and their modifications \citep{Schefzik2014} in order to provide practical guidance when to use which method. Ideally, the approaches of ECC and the Schaake shuffle or parts thereof could be combined, while retaining the advantages of each method. An initial attempt into this direction has been performed in \cite{Schefzik2014}, where a specific implemetation of the Schaake shuffle is proposed. In this approach, the empirical copula modeling the dependence is derived from past observations at dates for which the corresponding historical ensemble forecast resembles the current one with respect to a particular similarity criterion.

\appendix

\section{Appendix}\label{app}

\subsection{Proof of Theorem \ref{equiv}}\label{equiv.proof}

\noindent We first prove the implication from (1) to (2). If $i_\ell=0$ for at least one $\ell \in \{1,\ldots,L\}$, $D(\frac{i_1}{M},\ldots,\frac{i_L}{M})=0$ by axiom (D1) in Definition \ref{mdc}, in accordance with setting the empty sum equal to zero by convention, that is, $\sum_{n=1}^{0}c_n:=0$ for some sequence $(c_n)_{n \in \mathbb{N}}$. For $i_1,\ldots,i_L \in \{1,\ldots,M\}$, we set 
\begin{equation*}
a_{i_{1} \ldots i_{L}} := M \cdot \left(\Delta_{i_{L}-1}^{i_{L}} \cdots \Delta_{i_{1}-1}^{i_{1}}D\left(\frac{k_{1}}{M},\ldots,\frac{k_{L}}{M}\right)\right),
\end{equation*}
with $\Delta_{i_{\ell}-1}^{i_\ell}D$ as defined in axiom (D3) in Definition \ref{mdc}, and $A:=(a_{i_{1} \ldots i_{L}})_{i_{1},\ldots,i_{L}=1}^{M}$ and show that $A$ satisfies the axioms (A1) and (A2) from Definition \ref{stocharr} and therefore is an $L$-dimensional stochastic array of order $M$.
\begin{itemize}
\item[(A1)]{Since $D$ is a discrete copula, $D$ is $L$-increasing, that is,
\begin{equation*}
\Delta_{i_{L}-1}^{i_{L}} \cdots \Delta_{i_{1}-1}^{i_{1}}D \left(\frac{k_{1}}{M},\ldots,\frac{k_{L}}{M}\right) \geq 0. 
\end{equation*} 
Hence, $a_{i_{1} \ldots i_{L}} \geq 0$ by definition, and (A1) is fulfilled.}
\item[(A2)]{Now let $\ell \in \{1,\ldots,L\}$ be fixed. We have to show that 
\begin{equation*}
S_{1}:=\sum\limits_{i_{1}=1}^{M} \cdots \sum\limits_{i_{\ell-1}=1}^{M} \sum\limits_{i_{\ell+1}=1}^{M} \cdots \sum\limits_{i_{L}=1}^{M} a_{i_{1} \ldots i_{\ell-1} i_{\ell} i_{\ell+1} \ldots i_{L}} =1.
\end{equation*}
To this end, let $\lambda \in \{1,\ldots,L\} \setminus \{\ell\}$ and first consider $S_{2}:=\sum_{i_{\lambda}=1}^{M} a_{i_{1} \ldots i_{L}}$. By using the above definition of $a_{i_{1} \ldots i_{L}}$, writing down the sum $S_{2}$ explicitly yields that all of the $M \cdot 2^{L}$ addends $D(\cdot,\ldots,\cdot)$ of $S_{2}$ cancel out except for those $2^{L}$ having 0 or 1 in the $\lambda$-th component. Since discrete copulas are grounded according to axiom (D1) in Definition \ref{mdc}, all the $2^{L-1}$ terms of $S_{2}$ that have a 0 in the $\lambda$-th component vanish, and the $2^{L-1}$ terms 
\begin{equation*}
\begin{split}
S_{2}
= &M \cdot\\& \left(\Delta_{i_{L}-1}^{i_{L}} \cdots \Delta_{i_{\lambda+1}-1}^{i_{\lambda+1}} \Delta_{i_{\lambda-1}-1}^{i_{\lambda-1}} \cdots \Delta_{i_{1}-1}^{i_{1}} D \left(\frac{k_{1}}{M},\ldots,\frac{k_{\lambda-1}}{M},1,\frac{k_{\lambda+1}}{M},\ldots,\frac{k_{L}}{M} \right) \right)
\end{split}
\end{equation*}
remain. By writing down the multiple sum $S_{1}$ explicitly, iteratively applying the above considerations for the calculation of a sum of the type $S_2$ and accounting for the fact that discrete copulas are grounded due to axiom (D1) in Definition \ref{mdc}, all but two of the terms $D(\cdot,\ldots,\cdot)$ of $S_{1}$ vanish or cancel out, such that
\begin{eqnarray*}
S_{1}&=&M \cdot \left( \Delta_{i_{\ell}-1}^{i_{\ell}} D\left(1,\ldots,1,\frac{k_{\ell}}{M},1,\ldots,1 \right)\right)\\
&=&M \cdot \left(D \left(1,\ldots,1,\frac{i_{\ell}}{M},1,\ldots,1\right)-D \left(1,\ldots,1,\frac{i_{\ell}-1}{M},1,\ldots,1\right) \right)\\
&=&M \cdot \left(\frac{i_{\ell}}{M}-\frac{i_{\ell}-1}{M}\right)\\
&=&1,
\end{eqnarray*}
where the axiom (D2) in Definition \ref{mdc} is employed in the third equality. Hence, (A2) is fulfilled.}
\end{itemize}
Thus, $A$ is an $L$-dimensional stochastic array of order $M$.
\\
\\
Finally, the definition of $A$ gives the structure of $D$ in \eqref{stochm}. Indeed, similar arguments as in the proof of (A2) above yield that for fixed $\lambda \in \{1,\ldots,L\}$, the sum $S_3:=\sum_{\nu_\lambda=1}^{i_\lambda}a_{\nu_1 \ldots \nu_L}$ can be calculated as
\begin{equation*}
\begin{split}
S_3=& M \cdot \\ & \left(\Delta_{\nu_{L}-1}^{\nu_{L}} \cdots \Delta_{\nu_{\lambda+1}-1}^{\nu_{\lambda+1}} \Delta_{\nu_{\lambda-1}-1}^{\nu_{\lambda-1}} \cdots \Delta_{\nu_{1}-1}^{\nu_{1}} D \left(\frac{k_{1}}{M},\ldots,\frac{k_{\lambda-1}}{M},\frac{i_\lambda}{M},\frac{k_{\lambda+1}}{M},\ldots,\frac{k_{L}}{M} \right) \right).
\end{split}
\end{equation*}
For fixed $\ell \in \{1,\ldots,L\}$ and using the expression for $S_3$, similar calculations as in the proof of (A2) before yield
\begin{eqnarray*}
S_4&:=&\sum\limits_{\nu_1=1}^{i_1} \cdots \sum\limits_{\nu_{\ell-1}=1}^{i_{\ell-1}} \sum\limits_{\nu_{\ell+1}=1}^{i_{\ell+1}} \cdots \sum\limits_{\nu_L=1}^{i_L} a_{j_1 \ldots j_L}\\ &=& M \cdot \left(\Delta_{\nu_{\ell}-1}^{\nu_{\ell}} D \left(\frac{i_{1}}{M},\ldots,\frac{i_{\ell-1}}{M},\frac{k_\ell}{M},\frac{i_{\ell+1}}{M},\ldots,\frac{i_{L}}{M} \right) \right).
\end{eqnarray*}
We then employ $S_4$ to calculate
\begin{eqnarray*}
&&\frac{1}{M} \sum\limits_{\nu_{1}=1}^{i_{1}} \cdots \sum\limits_{\nu_\ell=1}^{i_\ell} \cdots \sum\limits_{\nu_{L}=1}^{i_{L}} a_{\nu_{1} \ldots \nu_{L}}=\frac{1}{M} \sum\limits_{\nu_\ell=1}^{i_\ell} \sum\limits_{\nu_{1}=1}^{i_{1}} \cdots \sum\limits_{\nu_{L}=1}^{i_{L}} a_{\nu_{1} \ldots \nu_{L}}\\
&=&\frac{1}{M} \sum\limits_{\nu_\ell=1}^{i_\ell} M \cdot \left(\Delta_{\nu_{\ell}-1}^{\nu_{\ell}} D \left(\frac{i_{1}}{M},\ldots,\frac{i_{\ell-1}}{M},\frac{k_\ell}{M},\frac{i_{\ell+1}}{M},\ldots,\frac{i_{L}}{M} \right) \right)\\
&=&\sum\limits_{\nu_\ell=1}^{i_\ell} \Delta_{\nu_{\ell}-1}^{\nu_{\ell}} D \left(\frac{i_{1}}{M},\ldots,\frac{i_{\ell-1}}{M},\frac{k_\ell}{M},\frac{i_{\ell+1}}{M},\ldots,\frac{i_{L}}{M} \right)\\
&=& \sum\limits_{\nu_\ell=1}^{i_\ell} \left[D \left(\frac{i_{1}}{M},\ldots,\frac{i_{\ell-1}}{M},\frac{\nu_\ell}{M},\frac{i_{\ell+1}}{M},\ldots,\frac{i_{L}}{M} \right)\right. \\&& \left. - D \left(\frac{i_{1}}{M},\ldots,\frac{i_{\ell-1}}{M},\frac{\nu_\ell-1}{M},\frac{i_{\ell+1}}{M},\ldots,\frac{i_{L}}{M} \right)\right]\\
&=&D \left(\frac{i_{1}}{M},\ldots,\frac{i_\ell}{M},\ldots,\frac{i_{L}}{M} \right)-\underbrace{D \left(\frac{i_{1}}{M},\ldots,\frac{1-1}{M},\ldots,\frac{i_{L}}{M} \right)}_{=0,\,\, \text{as $D$ is grounded}}\\
&=&D \left(\frac{i_{1}}{M},\ldots,\frac{i_{L}}{M} \right),
\end{eqnarray*}
where the second last equality takes advantage of a telescoping sum. Thus, \eqref{stochm} holds.
\\
\\
We now prove the implication from (2) to (1). Let the function $D$ be defined as in \eqref{stochm}. Obviously, $D$ has domain $\mbox{Dom}(D)=I_{M}^{L}$. Since $A$ is an $L$-dimensional stochastic array of order $M$, and according to the rules for multiple sums, we have  
\begin{eqnarray*}
0 &\leq& \frac{1}{M} \sum\limits_{\nu_{1}=1}^{i_{1}} \cdots \sum\limits_{\nu_{\ell}=1}^{i_{\ell}} \cdots \sum\limits_{\nu_{L}=1}^{i_{L}} a_{\nu_{1} \ldots \nu_{L}}= \frac{1}{M} \sum\limits_{\nu_{\ell}=1}^{i_{\ell}} \sum\limits_{\nu_{1}=1}^{i_{1}} \cdots \sum\limits_{\nu_{L}=1}^{i_{L}} a_{\nu_{1} \ldots \nu_{L}}\\ &\leq& \frac{1}{M} \sum\limits_{\nu_\ell=1}^{i_\ell} 1 = \frac{i_{\ell}}{M} \leq 1
\end{eqnarray*}
for fixed $i_{\ell} \in \{0,1,\ldots,M\}$, $\ell \in \{1,\ldots,L\}$, and hence get the range $\mbox{Ran}(D)=[0,1]$. Moreover, we have to check the axioms (D1), (D2) and (D3) in Definition \ref{mdc} for $D$.
\begin{itemize}
\item[(D1)]{Let $i_{\ell}=0$ for some $\ell \in \{1,\ldots,L\}$. Since the empty sum is equal to zero by convention, we get
\begin{equation*}
D\left(\frac{i_{1}}{M},\ldots,0,\ldots,\frac{i_{L}}{M}\right) = \frac{1}{M} \sum\limits_{\nu_{1}=1}^{i_{1}} \cdots \sum\limits_{\nu_{\ell}=1}^{0} \cdots \sum\limits_{\nu_{L}=1}^{i_{L}} a_{\nu_{1} \ldots \nu_{L}} = 0.
\end{equation*}
Clearly, this is also the case if there are two or more $\ell \in \{1,\ldots,L\}$ such that $i_{\ell}=0$. Hence, $D$ is grounded.
} 
\item[(D2)]{ Let $\ell \in \{1,\ldots,L\}$. Then,
\begin{eqnarray*}
D\left(1,\ldots,1,\frac{i_{\ell}}{M},1,\ldots,1 \right) 
&=& D\left(\frac{M}{M},\ldots,\frac{M}{M},\frac{i_{\ell}}{M},\frac{M}{M},\ldots,\frac{M}{M}\right)\\
&=&\frac{1}{M} \sum\limits_{\nu_{1}=1}^{M} \cdots \sum\limits_{\nu_{\ell-1}=1}^{M} \sum\limits_{\nu_{\ell}=1}^{i_{\ell}} \sum\limits_{\nu_{\ell+1}=1}^{M} \cdots \sum\limits_{\nu_{L}=1}^{M} a_{\nu_{1} \ldots \nu_{L}} \\
&=&\frac{1}{M} \sum\limits_{\nu_{\ell}=1}^{i_{ \ell}} \underbrace{\sum\limits_{\nu_{1}=1}^{M} \cdots \sum\limits_{\nu_{\ell-1}=1}^{M} \sum\limits_{\nu_{\ell+1}=1}^{M} \cdots \sum\limits_{\nu_{L}=1}^{M} a_{\nu_{1} \ldots \nu_{L}}}_{=1,\,\, \text{as $A$ is a stochastic array}} \\
&=&\frac{i_{\ell}}{M},
\end{eqnarray*}
according to the rules for multiple sums.
}
\item[(D3)]{We have to show that $D$ is $L$-increasing, that is,
\begin{equation*}
V:=\Delta_{i_{L}-1}^{i_{L}} \cdots \Delta_{i_{1}-1}^{i_{1}} D\left(\frac{k_{1}}{M},\ldots,\frac{k_{L}}{M}\right) \geq 0.
\end{equation*}
By definition, $V$ involves $2^{L}$ terms of the form $D(\cdot,\ldots,\cdot)$, where $2^{L-1}$ of them have positive sign and $2^{L-1}$ negative sign. Moreover, each of the $L$ arguments of a term $D(\cdot,\ldots,\cdot)$ is either of the form $\frac{i_{\ell}}{M}$ or of the form $\frac{i_{\ell}-1}{M}$ for $\ell \in \{1,\ldots,L\}$.
\\
\\
Let $\ell \in \{1,\ldots,L\}$ be fixed. In addition, let the arguments $\frac{k_{\lambda}}{M}$ for $\lambda \in \{1,\ldots,L\} \setminus \{\ell\}$ also be fixed, that is, $k_{\lambda}$ is either equal to $i_{\lambda}$ or equal to $i_{\lambda}-1$. First,
\begin{eqnarray}
&&D\left(\frac{k_{1}}{M},\ldots,\frac{i_{\ell}}{M},\ldots,\frac{k_{L}}{M}\right)-D\left(\frac{k_{1}}{M},\ldots,\frac{i_{\ell}-1}{M},\ldots,\frac{k_{L}}{M}\right)\nonumber\\
&=&\frac{1}{M} \sum\limits_{\nu_{1}=1}^{k_{1}} \cdots \sum\limits_{\nu_{\ell}=1}^{i_{\ell}} \cdots \sum\limits_{\nu_{L}=1}^{k_{L}} a_{\nu_{1} \ldots \nu_{L}} - \frac{1}{M} \sum\limits_{\nu_{1}=1}^{k_{1}} \cdots \sum\limits_{\nu_{\ell}=1}^{i_{\ell}-1} \cdots \sum\limits_{\nu_{L}=1}^{k_{L}} a_{\nu_{1} \ldots \nu_{L}}\label{msums}\\
&=& \frac{1}{M}\sum\limits_{\nu_{1}=1}^{k_{1}} \cdots \sum\limits_{\nu_{\ell}=i_{\ell}}^{i_{\ell}} \cdots \sum\limits_{\nu_{L}=1}^{k_{L}} a_{\nu_{1} \ldots \nu_{L}}\nonumber\\
&=& \frac{1}{M}\sum\limits_{\nu_{1}=1}^{k_{1}} \cdots \sum\limits_{\nu_{\ell-1}=1}^{k_{\ell-1}}\sum\limits_{\nu_{\ell+1}=1}^{k_{\ell+1}} \cdots \sum\limits_{\nu_{L}=1}^{k_{L}} a_{\nu_{1} \ldots \nu_{\ell-1}i_{\ell}\nu_{\ell+1} \ldots \nu_{L}},\nonumber
\end{eqnarray}
where to some extent, the multiple sum is now reduced due to the fact that the index $\nu_{\ell}=i_{\ell}$ is fixed.
\\
\\
Using the definition of $V$ and writing down $V$ explicitly in terms of the $2^{L}$ terms $D(\cdot,\ldots,\cdot)$ step by step, we obtain $2^{L-1}$ such differences as described above within the expression for $V$. Having calculated all those $2^{L-1}$ differences in the way as proposed before, we get $2^{L-2}$ new differences of the same type as in $\eqref{msums}$, where the index $\nu_{1}=i_{1}$ in the multiple sums becomes fixed, and can thus proceed as before. By applying this scheme successively, we finally end up with
\begin{equation*}
V = \frac{1}{M}\sum\limits_{\nu_{L}=1}^{i_{L}}a_{i_{1}i_{2} \ldots i_{L-1}\nu_{L}}-\frac{1}{M}\sum\limits_{\nu_{L}=1}^{i_{L}-1}a_{i_{1}i_{2} \ldots i_{L-1}\nu_{L}}=\frac{1}{M}\, a_{i_{1}i_{2} \ldots i_{L-1}i_{L}}.
\end{equation*}
Since $A=(a_{i_{1} \ldots i_{L}})_{i_{1} \ldots i_{L}}^{M}$ is an $L$-dimensional stochastic array of order $M$ by assumption, $a_{i_{1}i_{2} \ldots i_{L-1}i_{L}} \geq 0$ for all $i_{\ell} \in \{1,\ldots,M\}$, $\ell \in \{1,\ldots,L\}$. Hence, $V \geq 0$, and $D$ is $L$-increasing.
}
\end{itemize}
Thus, $D$ is indeed a discrete copula.$\hfill\qed$

\subsection{Proof of Lemma \ref{extension}}\label{extension.proof}

Let 
\begin{equation*}
J_{M}^{(\ell)} := \left\{0=:\frac{b_{0}^{(\ell)}}{M} < \frac{b_{1}^{(\ell)}}{M} < \cdots < \frac{b_{r_{\ell}}^{(\ell)}}{M} < \frac{b_{r_{\ell}+1}^{(\ell)}}{M}:=1 \right\}
\end{equation*}
for $\ell \in \{1,\ldots,L\}$, with the corresponding equivalent sets
\begin{equation*}
K_{M}^{(\ell)} := \{0=:a_{0}^{(\ell)} < a_{1}^{(\ell)} < \cdots < a_{r_{\ell}}^{(\ell)} <a_{r_{\ell}+1}^{(\ell)}:=M \}.
\end{equation*}
According to Theorem \ref{equiv}, it suffices to construct an $L$-dimensional permutation array $A$ of order $M$ to get an irreducible discrete extension copula $D$ of an irreducible discrete subcopula $D^{\ast}$. The array $A$ has to be such that each block specified by the positions $(a_{s_{1}}^{(1)},a_{s_{2}}^{(2)},\ldots,a_{s_{L}}^{(L)})$ and $(a_{s_{1}+1}^{(1)},a_{s_{2}+1}^{(2)},\ldots,a_{s_{L}+1}^{(L)})$, which consists of the rows from 
$a_{s_{1}}^{(1)}+1$ to $a_{s_{1}+1}^{(1)}$, from $a_{s_{2}}^{(2)}+1$ to $a_{s_{2}+1}^{(2)}$, and so forth, up to the row from $a_{s_{L}}^{(L)}+1$ to $a_{s_{L}+1}^{(L)}$, contains a number of 1's equal to the volume
\begin{equation*}
M \cdot \left(\Delta_{a_{s_{L}}^{(L)}}^{a_{s_{L}+1}^{(L)}} \cdots \Delta_{a_{s_{1}}^{(1)}}^{a_{s_{1}+1}^{(1)}} D^{\ast} \left(\frac{k_{1}}{M},\ldots,\frac{k_{L}}{M}\right)\right),
\end{equation*}
where $s_{\ell} \in \{0,\ldots,r_{\ell}\}$ and $\ell \in \{1,\ldots,L\}$.
\\
\\
To show the existence of such a permutation array $A$, let $\ell \in \{1,\ldots,L\}$ be fixed and consider
the subarray which contains all the blocks determined by the positions $(a_{s_{1}}^{(1)},\ldots,a_{s_{L}}^{(L)})$ and $(a_{s_{1}+1}^{(1)},\ldots,a_{s_{L}+1}^{(L)})$ for all $s_{\lambda} \in \{0,\ldots,r_{\lambda}\}$, where $\lambda \in \{1,\ldots,L\} \setminus \{\ell\}$.
\\
\\
We need to show that the number $a_{s_{\ell}+1}^{(\ell)} - a_{s_{\ell}}^{(\ell)}$ of rows in this subarray is equal to the number of 1's corresponding to all those blocks. This indeed holds, as
\begin{eqnarray*}
\sum\limits_{s_{1}=0}^{r_{1}} &\cdots& \sum\limits_{s_{\ell-1}=0}^{r_{\ell-1}}\sum\limits_{s_{\ell+1}=0}^{r_{\ell+1}} \cdots \sum\limits_{s_{L}=0}^{r_{L}} M \cdot \left(\Delta_{a_{s_{L}}^{(L)}}^{a_{s_{L}+1}^{(L)}} \cdots \Delta_{a_{s_{1}}^{(1)}}^{a_{s_{1}+1}^{(1)}} D^{\ast} \left(\frac{k_{1}}{M},\ldots,\frac{k_{L}}{M}\right)\right)\nonumber\\
&=&M \cdot \sum\limits_{s_{1}=0}^{r_{1}} \cdots \sum\limits_{s_{\ell-1}=0}^{r_{\ell-1}}\sum\limits_{s_{\ell+1}=0}^{r_{\ell+1}} \cdots \sum\limits_{s_{L}=0}^{r_{L}}  \left(\Delta_{a_{s_{L}}^{(L)}}^{a_{s_{L}+1}^{(L)}} \cdots \Delta_{a_{s_{1}}^{(1)}}^{a_{s_{1}+1}^{(1)}} D^{\ast} \left(\frac{k_{1}}{M},\ldots,\frac{k_{L}}{M}\right)\right)\nonumber \\
&=& M \cdot \left( \Delta_{a_{s_{\ell}}^{(\ell)}}^{a_{s_{\ell}+1}^{(\ell)}} D^{\ast}\left(1,\ldots,1,\frac{k_{\ell}}{M},1,\ldots,1 \right)\right)\nonumber\\
&=& M \cdot \left(D^{\ast} \left(1,\ldots,1,\frac{a_{s_{\ell}+1}^{(\ell)}}{M},1,\ldots,1 \right)- D^{\ast} \left(1,\ldots,1,\frac{a_{s_{\ell}}^{(\ell)}}{M},1,\ldots,1 \right) \right)\nonumber\\
&=& M \cdot \left(\frac{a_{s_{\ell}+1}^{(\ell)}}{M}-\frac{a_{s_{\ell}}^{(\ell)}}{M} \right)\nonumber\\
&=&a_{s_{\ell}+1}^{(\ell)}-a_{s_{\ell}}^{(\ell)},
\end{eqnarray*}
\noindent where we use axiom (S2) in Definition \ref{mdsubcop} for the second last equality.
\\
\\
To see the second equality explicitly, we proceed analogously as in the proof of axiom (A2) in Theorem \ref{equiv} for the implication from (1) to (2). We set
\begin{equation*}
S:=\sum\limits_{s_{1}=0}^{r_{1}} \cdots \sum\limits_{s_{\ell-1}=0}^{r_{\ell-1}}\sum\limits_{s_{\ell+1}=0}^{r_{\ell+1}} \cdots \sum\limits_{s_{L}=0}^{r_{L}}  \left(\Delta_{a_{s_{L}}^{(L)}}^{a_{s_{L}+1}^{(L)}} \cdots \Delta_{a_{s_{1}}^{(1)}}^{a_{s_{1}+1}^{(1)}} D^{\ast} \left(\frac{k_{1}}{M},\ldots,\frac{k_{L}}{M}\right)\right),
\end{equation*}
let $\lambda \in \{1,\ldots,L\} \setminus \{\ell\}$ be fixed and first consider the sum
\begin{equation*}
T:=\sum\limits_{s_\lambda=0}^{r_\lambda} \left(\Delta_{a_{s_{L}}^{(L)}}^{a_{s_{L}+1}^{(L)}} \cdots \Delta_{a_{s_{1}}^{(1)}}^{a_{s_{1}+1}^{(1)}} D^{\ast} \left(\frac{k_1}{M},\ldots,\frac{k_L}{M} \right) \right).
\end{equation*}
The $(r_{\lambda}+1) \cdot 2^L$ addends $D(\cdot,\ldots,\cdot)$ of $T$ cancel except for those $2^L$ having a 0 or a 1 in the $\lambda$-th component, which indeed occurs as $a_{0}^{(\lambda)}=0$ and $a_{r_{\lambda}+1}^{(\lambda)}=M$. According to axiom (S1) in Definition \ref{mdsubcop}, all the $2^{L-1}$ terms having a 0 in the $\lambda$-th component vanish, and we obtain
\begin{equation*}
T=\Delta_{a_{s_{L}}^{(L)}}^{a_{s_{L}+1}^{(L)}} \cdots \Delta_{a_{s_{\lambda+1}}^{(\lambda+1)}}^{a_{s_{\lambda+1}+1}^{(\lambda+1)}} \Delta_{a_{s_{\lambda-1}}^{(\lambda-1)}}^{a_{s_{\lambda-1}+1}^{(\lambda-1)}} \cdots \Delta_{a_{s_{1}}^{(1)}}^{a_{s_{1}+1}^{(1)}} D^{\ast} \left(\frac{k_1}{M},\ldots,\frac{k_{\lambda-1}}{M},1,\frac{k_{\lambda+1}}{M},\ldots,\frac{k_L}{M} \right).
\end{equation*}
Applying this iteratively and using again axiom (S1) in Definition \ref{mdsubcop}, all but two of the terms $D(\cdot,\ldots,\cdot)$ of $S$ vanish or cancel out, such that
\begin{equation*}
S=\Delta_{a_{s_{\ell}}^{(\ell)}}^{a_{s_{\ell}+1}^{(\ell)}} D^{\ast} \left(1,\ldots,1,\frac{k_\ell}{M},1,\ldots,1 \right),
\end{equation*}
as desired.$\hfill\qed$

\subsection{Proof of Theorem \ref{sklarmdc}}\label{sklarmdc.proof}

\begin{enumerate}
\item{This is just a special case of the common Sklar's theorem. The claim follows straightforwardly by checking the well-known axioms of a finite $L$-dimensional CDF for $H$ as defined in \eqref{sklar}.}
\item{Let $H$ be a finite $L$-dimensional CDF with $\mbox{Ran}(H)\subseteq I_{M}$ having univariate marginal CDFs $F_{1},\ldots,F_{L}$. Set
\begin{equation*}
J_{M}^{(\ell)} := \left\{\frac{i_{\ell}}{M} \in I_{M}\left| \frac{i_{\ell}}{M} \in \mbox{Ran}(F_{\ell})\right\} \supseteq \{0,1\}\right.
\end{equation*}
for $\ell \in \{1,\ldots,L\}$ and define
\begin{equation*}
D^{\ast}:J_{M}^{(1)} \times \cdots \times J_{M}^{(L)} \rightarrow I_{M}, \,\,\, D^{\ast}\left(\frac{i_{1}}{M},\ldots,\frac{i_{L}}{M}\right):=H(y_{1},\ldots,y_{L}),
\end{equation*}
where $y_{\ell}$ satisfies $F_{\ell}(y_{\ell})=\frac{i_{\ell}}{M}$ for $\ell \in \{1,\ldots,L\}$.
\\
\\
We now show that $D^{\ast}$ is an irreducible discrete subcopula. First, $\mbox{Ran}(H) \subseteq I_{M}$ by assumption, and $D^{\ast}$ is well-defined, due to the well-known fact that $H(y_1,\ldots,y_L)=H(z_1,\ldots,z_L)$ for points $y_1,\ldots,y_L \in \overline{\mathbb{R}}$ and $z_1,\ldots,z_L \in \overline{\mathbb{R}}$ such that $F(y_\ell)=F(z_\ell)$ for all $\ell \in \{1,\ldots,L\}$. Furthermore, the axioms (S1), (S2) and (S3) for discrete subcopulas in Definition \ref{mdsubcop} are fulfilled, as shown in what follows.
\begin{itemize}
\item[(S1)]{Let $i_{\ell}=0$ for an $\ell \in \{1,\ldots,L\}$. Then,
\begin{equation*} 
D^{\ast}\left(\frac{i_{1}}{M},\ldots,\frac{i_{\ell-1}}{M},0,\frac{i_{\ell+1}}{M},\ldots,\frac{i_{L}}{M}\right)=H(y_{1},\ldots,y_{L})
\end{equation*} 
with $F_{\ell}(y_{\ell})=\frac{0}{M}=0$ and $F_{\lambda}(y_{\lambda})=\frac{i_{\lambda}}{M}$ for all $\lambda \in \{1,\ldots,L\} \setminus \{\ell\}$. However,
\begin{equation*}
F_{\ell}(y_{\ell}) = H(\infty,\ldots,\infty,y_{\ell},\infty,\ldots,\infty) = 0,
\end{equation*} 
and since $H$ is non-decreasing in each argument, we have 
\begin{equation*}
H(y_{1},\ldots,y_{\ell},\ldots,y_{L})=0,
\end{equation*}
and hence
\begin{equation*}
D^{\ast}\left(\frac{i_{1}}{M},\ldots,\frac{i_{\ell-1}}{M},0,\frac{i_{\ell+1}}{M},\ldots,\frac{i_{L}}{M}\right) = 0
\end{equation*} 
for all $\frac{i_{\lambda}}{M} \in J_{M}^{(\lambda)}$, where $\lambda \in \{1,\ldots,L\} \setminus \{\ell\}$. Clearly, this is also true if $i_{\ell}=0$ for two or more $\ell \in \{1,\ldots,L\}$.}
\item[(S2)]{For $\ell \in \{1,\ldots,L\}$, consider
\begin{equation*}
D^{\ast}\left(1,\ldots,1,\frac{i_{\ell}}{M},1,\ldots,1\right)=H(y_{1},\ldots,y_{L})
\end{equation*}
with $F_{\ell}(y_{\ell})=\frac{i_{\ell}}{M}$ and $F_{\lambda}(y_{\lambda})=1$ for $\lambda \in \{1,\ldots,L\} \setminus \{\ell\}$. Set $y_{\lambda}:=\infty$ for $\lambda \in \{1,\ldots,L\} \setminus \{\ell\}$. Then,
\begin{equation*}
D^{\ast}\left(1,\ldots,1,\frac{i_{\ell}}{M},1,\ldots,1\right)=H(\infty,\ldots,\infty,y_{\ell},\infty,\ldots,\infty)=F_{\ell}(y_{\ell})=\frac{i_{\ell}}{M}
\end{equation*}  
for all $\frac{i_{\ell}}{M} \in J_{M}^{(\ell)}$.}
\item[(S3)]{To show that $D^{\ast}$ is $L$-increasing, we use the $L$-increasingness of $H$ as a multivariate CDF and obtain
\begin{equation*}
\Delta_{i_{L}}^{j_{L}} \cdots \Delta_{i_{1}}^{j_{1}} D^{\ast}\left(\frac{k_{1}}{M},\ldots,\frac{k_{L}}{M}\right)=\Delta_{y_L}^{z_{L}} \cdots \Delta_{y_{1}}^{z_{1}} H(x_{1},\ldots,x_{L}) \geq 0
\end{equation*}
for all $\frac{i_{\ell}}{M},\frac{j_{\ell}}{M} \in J_{M}^{(\ell)}$ such that $\frac{i_{\ell}}{M} \leq \frac{j_{\ell}}{M}$ , where $F_{\ell}(y_{\ell})=\frac{i_{\ell}}{M}$ and $F_{\ell}(z_{\ell})=\frac{j_{\ell}}{M}$ for all $y_{\ell},z_{\ell} \in \overline{\mathbb{R}}$ and $\ell \in \{1,\ldots,L\}$. Hence, $D^{\ast}$ is $L$-increasing.
}
\end{itemize}
Thus, $D^{\ast}$ is indeed a subcopula.
\\
\\
According to Lemma \ref{extension}, $D^{\ast}$ can be extended to a discrete copula $D$, which satisfies
\begin{eqnarray*}
D(F_{1}(y_{1}),\ldots,F_{L}(y_{L})) &=& D\left(\frac{i_{1}}{M},\ldots,\frac{i_{L}}{M}\right) = D^{\ast} \left(\frac{i_{1}}{M},\ldots,\frac{i_{L}}{M}\right)\\ &=& H(y_{1},\ldots,y_{L})
\end{eqnarray*}
for $y_{1},\ldots,y_{L} \in \overline{\mathbb{R}}$. Hence, $H(y_{1},\ldots,y_{L})=D(F_{1}(y_{1}),\ldots,F_{L}(y_{L}))$.
\\
\\
If $\mbox{Ran}(F_{\ell})=I_{M}$ for all $\ell \in \{1,\ldots,L\}$, then the discrete subcopula $D^{\ast}$ has domain $I_{M}^{L}$, and thus we have $D=D^{\ast}$, that is, $D$ is uniquely determined.$\hfill\qed$}
\end{enumerate}

%
\section*{Acknowledgments}
I gratefully acknowledge support by the Klaus Tschira Foundation and the VolkswagenStiftung under the project 
``Mesoscale Weather Extremes: Theory, Spatial Modeling and Prediction'' and by Deutsche Forschungsgemeinschaft
through the Research Training Group RTG 1953.\\
Most of the work has been done during my time as a PhD student at the Institute for Applied Mathematics at Heidelberg University. I thank my advisors Tilmann Gneiting and Thordis Thorarinsdottir for many helpful discussions.\\
Moreover, I thank an anonymous reviewer of this paper, who has provided many valuable comments and suggestions.\\
For the real data illustration in Figure \ref{ecc}, the forecasts have been made available by the European Center for Medium-Range Weather Forecasts and the observations by the German Weather Service, which is gratefully acknowledged.

%






%
%
%

\end{document}